\documentclass[a4paper,11pt]{article}
\usepackage{jcappub} 
\usepackage{lineno}
\usepackage{orcidlink} 
\usepackage{xspace}
\usepackage[dvipsnames]{xcolor}
\usepackage{float}
\usepackage{subcaption}
\usepackage[percent]{overpic}
\usepackage[table,dvipsnames]{xcolor}
\usepackage{tabularx}
\usepackage{multirow}
\usepackage{booktabs}
\usepackage[table]{xcolor}
\usepackage{amsmath} 
\usepackage{array}
\usepackage{colortbl}
\usetikzlibrary{shapes,arrows,positioning,calc}
\usepackage{xcolor}

\usetikzlibrary{shapes.geometric, arrows, positioning, fit}
\definecolor{lightgray}{gray}{0.93}
\definecolor{headergray}{gray}{0.82}
\usetikzlibrary{shadows, arrows.meta}
\definecolor{deepblue}{RGB}{25,100,180}
\definecolor{deeporange}{RGB}{230,110,30}
\definecolor{lightblue}{RGB}{210,230,255}
\definecolor{lightorange}{RGB}{255,230,210}
\newcolumntype{M}[1]{>{\centering\arraybackslash}m{#1}}

\usepackage{tikz}
\usetikzlibrary{
  calc,              
  arrows.meta, 
  positioning, 
  shapes.geometric
}

\usetikzlibrary{
  matrix,          
  calc,            
  arrows.meta,     
  positioning,     
  shapes.geometric 
}

\definecolor{crisp}{RGB}{200, 200, 200}
\colorlet{Mycolor1}{green!10!orange}
\usepackage{tcolorbox}
\tcbuselibrary{theorems, skins}

\usepackage{mdframed}

\title{\boldmath KGB-evolution: a relativistic $N$-body code for kinetic gravity braiding models}

\newcommand{\hiclass}{\texttt{hi\_class}\xspace}
\newcommand{\class}{\texttt{CLASS}\xspace}

\newcommand{\gev}{\texttt{gevolution}\xspace}
\newcommand{\kev}{\texttt{k-evolution}\xspace}
\newcommand{\kgb}{\texttt{KGB-evolution}\xspace}

\author[a]{Ahmad Nouri-Zonoz \orcidlink{0009-0006-6164-8670},}
\affiliation[a]{D\'epartement de Physique Th\'eorique and Centre for Astroparticle Physics, Universit\'e de Gen\`eve,
1211 Gen\`eve 4 , Switzerland}
\author[b]{Farbod Hassani \orcidlink{0000-0003-2640-4460},}
\affiliation[b]{Institute of Theoretical Astrophysics, University of Oslo,
0315 Oslo, Norway}
\author[c]{Emilio Bellini \orcidlink{0000-0003-4762-0795},}
\affiliation[c]{\textit{Center for Astrophysics and Cosmology}, University of Nova Gorica, Nova Gorica, Slovenia}
\author[a]{and Martin Kunz \orcidlink{0000-0002-3052-7394},}

\emailAdd{Ahmadreza.Nourizonoz@unige.ch, farbod.hassani@astro.uio.no, emilio.bellini@ung.si, martin.kunz@unige.ch}

\abstract{We present \kgb, a relativistic $N$-body simulation code that extends the \kev code by incorporating an effective field theory parameterization of kinetic gravity braiding, while also including the $k$-essence model as a limiting case.
As a first step, we implement the linearized dark energy stress–energy tensor and scalar field equations, providing the groundwork for a future full Horndeski theory extension. We validate \kgb by comparing its  power spectra against linear predictions from \hiclass, finding excellent agreement on large scales at low redshifts and over all scales at high redshifts. We demonstrate that nonlinear growth of matter and metric perturbations on small scales drives the linearized dark energy field into a nonlinear clustering regime, which in turn feeds back on the growth of cosmic structure. In contrast to the $k$‑essence limit, a nonzero braiding  considerably amplifies this backreaction, producing a significantly stronger alteration of structure formation in the kinetic gravity braiding model.}

\begin{document}
\maketitle
\flushbottom

\definecolor{lightblue}{RGB}{173,216,230}
\definecolor{lightgreen}{RGB}{144,238,144}
\definecolor{lightyellow}{RGB}{255,255,224}
\definecolor{lightred}{RGB}{255,182,193}

\section{Introduction}
\label{sec:intro}
The nature of cosmic acceleration still remains one of the most significant unresolved questions in cosmology. While the standard cosmological constant ($\Lambda$) has successfully explained the past observations \citep{Planck:2015bpv, BOSS:2016wmc}, recent high precision measurements, particularly from Dark Energy Spectroscopic Instrument (DESI)\footnote{\url{https://www.desi.lbl.gov/}}, provide evidence for the dynamical evolution of dark energy over the history of the universe \cite{DESI:2025zgx}. In particular, analyses based on DESI DR2  indicate that the equation of state parameter for dark energy ($w$) may have crossed the so-called \textit{phantom divide} line $(w = -1)$ \citep{DESI:2025fii}.

A theoretical framework capable of capturing such dynamical behavior and potentially providing a more robust alternative to the $\Lambda$ is the Horndeski theory \cite{Horndeski:1974wa}, which is among the most general scalar–tensor theories with second-order equations of motion, and serves as the foundation for later extensions such as beyond-Horndeski \cite{Gleyzes:2014dya} and DHOST \cite{Langlois:2015cwa}. Among the subclasses of Horndeski theories, the simplest and most popular one is known as $k$-essence \cite{PhysRevLett.85.4438, PhysRevD.62.023511}, which involves a minimally coupled scalar field whose dynamics are governed solely by a non-canonical scalar field term in the Lagrangian. This theory is implemented in several Boltzmann solvers, such as \class \citep{Lesgourgues:2011re, Blas:2011rf}, where dark energy is treated using a fluid approximation, and \hiclass \citep{Zumalacarregui:2016pph}, which directly solves the scalar field dynamics. It is also incorporated in $N$-body simulations, such as \kev \citep{Hassani:2019lmy, Hassani:2019wed}, using the Effective Field Theory (EFT) formalism. Despite their simplicity, standard $k$-essence models, generally fail to cross the phantom divide due to constraints from their evolution equation and the presence of instabilities, see for example \cite{Vikman:2004dc,Kunz:2006wc,Creminelli:2008wc,Zhao:2005vj}. On the other hand, a more complex model within the Horndeski framework known as kinetic gravity braiding (KGB) is able to cross the phantom barrier. This class still features a minimally coupled scalar field but introduces a distinctive interaction, known as \textit{braiding}, between the scalar field's kinetic term and the metric. This mixing of scalar and tensor kinetic terms affects the evolution of perturbations and allows crossing the phantom divide without introducing ghost and gradient instabilities \cite{Deffayet:2010qz}. 

In this work, we study the linear evolution of the KGB model in the presence of nonlinear dark matter clustering.  The full nonlinear treatment of dark energy self‑interactions — which might give rise to nonlinear instabilities in the scalar sector (see \cite{Hassani:2021tdd, Hassani:2022xyb, Eckmann:2022wtd, Sawicki:2012pz}) — will be pursued in a dedicated follow up study. To implement the KGB model, we choose \gev \citep{Adamek:2015eda, Adamek:2016zes}, a relativistic $N$-body simulation code designed for cosmic structure formation under the weak field approximation of general relativity. The modular structure of \gev provides a flexible platform for including dark energy and modified gravity extensions without altering the core solver. Furthermore, to derive the KGB equations we follow the $\alpha$-parametrization formalism introduced in \citep{Bellini:2014fua}.

The paper is structured as follows: In Sec.\ \ref{sec:EQsofKGB}, we begin with the full covariant action and derive both the modified linearized Einstein field equations and the scalar field equation of motion for dark energy. In Sec.\ \ref{sec:implementation}, we explain the implementation of these linearized equations within the \kev/\gev framework and compare the perturbation treatment schemes used in different codes. We also introduce the 
scalar field and numerical configuration options added to the \gev settings. In Sec.\ \ref{sec:results} we analyze the power spectra of different fields in both the $k$-essence and KGB models and perform consistency checks through comparing $N$-body results with \hiclass. We then explore the nonlinear clustering of dark energy via both power spectra and 2D/3D snapshot analyses and quantify the 
effect of braiding on structure growth. Finally, Sec.\ \ref{sec:conclusions} summarizes our main results and outlines future directions.

\section{Horndeski Formulation of Kinetic Gravity Braiding}
\label{sec:EQsofKGB}
Horndeski theory is described by the following Lagrangian density
\begin{equation}
     \mathcal{L}_{\text{H}} = \sum_{i=2}^{5} \mathcal{L}_i \, , 
\end{equation}
where $\mathcal{L}_i$ are given by\footnote{This form of the Horndeski Lagrangian was first introduced in \cite{Kobayashi:2011nu}}
\begin{align*}
\mathcal{L}_2 &= G_2(\phi, X) \, ,\\
\mathcal{L}_3 &= -G_3(\phi, X) \Box \phi \, ,\\
\mathcal{L}_4 &= G_4(\phi, X) R + G_{4X} \left[ (\Box \phi)^2 - (\nabla_\mu \nabla_\nu \phi)^2 \right] \, ,\\
\mathcal{L}_5 &= G_5(\phi, X) G_{\mu\nu} \nabla^\mu \nabla^\nu \phi - \frac{1}{6} G_{5X} \left[ (\Box \phi)^3 - 3\Box \phi (\nabla_\mu \nabla_\nu \phi)^2 + 2(\nabla_\mu \nabla_\nu \phi)^3 \right]  \, .
\end{align*}
Here, $G_i$ and their derivatives $G_{iX}$ are arbitrary functions of the scalar field $\phi$ and its kinetic term $X = -\frac{1}{2}\nabla^\mu\phi\,\nabla_\mu\phi$. Within this general framework, the $k$-essence and KGB models, together constitute a particular subclass by choosing $G_5 = 0$ 
 and fixing $G_4=M_{\rm Pl}^2/2$, which reproduces the standard Einstein–Hilbert term. In this case, the dynamics are then entirely encoded in $G_2$ and $G_3$ functions.
 
 For clarity, it is often convenient to separate the Einstein–Hilbert piece from the scalar sector. The total action can then be written as
\begin{equation}
    \mathcal{S}_{\rm tot} = \int d^4x \sqrt{-g} \left(  \frac{M_{\rm Pl}^2 R}{2}+ \mathcal{L}_{k\text{-essence}}+ \mathcal{L}_{\rm KGB}+\mathcal{L}_{\rm m} \right) \, ,
	\label{eq:action}
\end{equation}
where $M_{\rm Pl}^2 = \frac{1}{8 \pi G}$ is the squared Planck mass, $\mathcal{L}_{k\text{-essence}} = G_2(\phi, X)$ and $\mathcal{L}_{\rm KGB} = -G_3(\phi, X) \Box \phi$. 
Furthermore, $\mathcal{L}_{\rm m}$ is the matter Lagrangian. 
For cold dark matter (CDM) particles in \gev, it is modeled as a collection 
of massive, non-relativistic point particles—equivalently, a sum over 
single-particle Lagrangians \citep{Adamek:2016zes}.

To study inhomogeneities in the Universe we adopt the perturbed Friedmann-Lema\^itre-Robertson-Walker (FLRW) metric in the Poisson gauge which reads as
\begin{equation}
d s^2=a^2(\tau)\left[-e^{2 \Psi} d \tau^2-2 B_i d x^i d t+\left(e^{-2 \Phi} \delta_{i j}+h_{i j}\right) d x^i d x^j\right] \, ,
	\label{eq:FLRW}
\end{equation}
where $a(\tau)$ is the scale factor, $d\tau = dt/a(t)$ is the differential element of conformal time, $\Psi$ and $\Phi$ denote the temporal and spatial scalar perturbations respectively, and $B_i$ and $h_{ij}$ represent the transverse vector ($\partial^i B_i = 0$) and transverse-traceless tensor ($\partial^i h^j_i = h^i_i = 0$) perturbations. We also adopt the Einstein summation convention for all repeated indices.

\subsection{Modified Einstein field equations and the scalar field evolution}
The modified Einstein field equations are obtained by taking the variation of the action \eqref{eq:action} with respect to the metric
\begin{equation}
\begin{aligned}
\frac{-2}{\sqrt{-g}}\frac{\delta\mathcal{{S}}_\text{tot}}{\delta g^{\mu\nu}}&=  M_{\text{Pl}}^2 R_{\mu \nu} - \frac{1}{2} M_{\text{Pl}}^2 R g_{\mu \nu}  - g_{\mu \nu}G_2
- \nabla_{\mu} \phi \nabla_{\nu} \phi G_{2X}- g_{\mu \nu} \nabla_{\alpha} X \nabla^{\alpha} \phi G_{3X}\\
& \quad  + \nabla_{\gamma} \nabla^\gamma \phi \nabla_{\mu} \phi \nabla_{\nu} \phi G_{3X} + \nabla_{\mu} X \nabla_{\nu} \phi G_{3X}  
+ \nabla_{\mu} \phi \nabla_{\nu} X G_{3X} - g_{\mu \nu} \nabla_{\alpha} \phi \nabla^\alpha \phi G_{3\phi} \\
& \quad+ 2 \nabla_{\mu} \phi \nabla_{\nu} \phi G_{3\phi} - T^{(\rm m)}_{\mu\nu} \, ,
\end{aligned}
\label{varLtot2}
\end{equation}
where the arguments of the functions $G_i(\phi, X)$ are omitted for simplicity. Moreover, $T^{(\mathrm m)}_{\mu\nu}$ corresponds to the variation of the matter Lagrangian with respect to the metric.
The reader can find the explicit particle-based expressions for the matter stress–energy tensor in \cite{Adamek:2016zes}. In our implementation, we leave the matter sector in \gev unchanged and focus only on modifying the dark energy equations.
Therefore, from \eqref{varLtot2}, we can write the modified Einstein field equations in the following form 
\begin{equation}
M_{\text{Pl}}^2 G_{\mu \nu} = T^{ (\rm m)}_{\mu \nu} + T^{(\phi)}_{\mu \nu} \, ,
\label{ModifiedEinstein}
\end{equation}
where
\begin{align}
G_{\mu \nu} &= R_{\mu\nu}-\frac{1}{2}Rg_{\mu\nu} \, , \\
    T^{(\phi)}_{\mu \nu}&=  g_{\mu \nu} G_2 +  \nabla_{\mu} \phi \nabla_{\nu} \phi G_{2X} 
      +g_{\mu \nu} \nabla_{\alpha} X \nabla^\alpha \phi G_{3X} - \nabla_{\alpha} \nabla^\alpha \phi \nabla_{\mu} \phi \nabla_{\nu} \phi G_{3X}\nonumber\\
     & \quad - \nabla_{\mu} X \nabla_{\nu} \phi G_{3X}- \nabla_{\mu} \phi \nabla_{\nu} X G_{3X}+ g_{\mu \nu} \nabla_{\alpha} \phi \nabla^\alpha \phi G_{3\phi}- 2 \nabla_{\mu} \phi \nabla_{\nu} \phi G_{3\phi} \, .
    \label{scalarstress}
\end{align}

Similarly, we can derive the equation of motion for 
dark energy scalar field 
by taking the variation of the action \eqref{eq:action}  with respect to the scalar field
\begin{equation}
\begin{aligned}
\frac{-2}{\sqrt{-g}}\frac{\delta\mathcal{{S}}_\text{tot}}{\delta \phi}&=-2 \nabla^{\alpha  }\nabla_{\alpha  }\phi  G_{{2}{}X} + 2 \nabla_{\alpha  }\nabla_{\beta  }\nabla^{\beta  }\phi  \nabla^{\alpha  \
}\phi  G_{{3}{}X}  -2 \nabla_{\alpha  }\nabla_{\beta  }\phi  \nabla^{\alpha  }\nabla^{\beta  }\phi  G_{{3}{}X} \\
&~\quad +2 \nabla_{\alpha  }\
\nabla^{\alpha  }\phi  \nabla^{\beta  }\nabla_{\beta  }\phi  \
G_{{3}{}X}  -2 \nabla_{\alpha  }\phi  \nabla^{\beta  }\nabla_{\beta  \
}\nabla^{\alpha  }\phi  G_{{3}{}X} + 2 \nabla_{\alpha  }\nabla^{\beta \
 }\phi  \nabla^{\alpha  }\phi  \nabla_{\beta  }\phi  G_{{2}{}X X} \\
 &~\quad-2 \nabla_{\alpha  }\nabla^{\beta  }\phi  \nabla^{\alpha  }\phi  \nabla_{\beta  }\phi  \nabla_{\gamma  }\nabla^{\gamma  }\phi  G_{{3}{}X \
X} + 2 \nabla_{\alpha  }\phi  \nabla_{\beta  }\phi  \nabla_{\gamma  }\
\nabla^{\beta  }\phi  \nabla^{\gamma  }\nabla^{\alpha  }\phi  \
G_{{3}{}X X} \\
&~\quad -2 G_{{2}{}\phi } + 2 \nabla_{\alpha  }\nabla^{\alpha  \
}\phi  G_{{3}{}\phi } + 2 \nabla^{\alpha  }\nabla_{\alpha  }\phi  \
G_{{3}{}\phi }  -2 \nabla_{\alpha  }\phi  \nabla^{\alpha  }\phi  \
G_{{2}{}\phi  X} \\
&~\quad -4 \nabla_{\alpha  }\nabla^{\beta  }\phi  \nabla^{\
\alpha  }\phi  \nabla_{\beta  }\phi  G_{{3}{}\phi  X} + 2 \nabla_{\alpha  }\phi  \nabla^{\alpha  }\phi  \nabla_{\beta  }\nabla^{\beta  }\phi  G_{{3}{}\phi  X} + 2 \nabla_{\alpha  }\phi  \nabla^{\alpha  }\phi  G_{{3}{}\phi  \phi } \, .
\end{aligned}
\label{scalarEOM}
\end{equation}

To proceed further, it is useful to recast the scalar field contribution into an effective fluid description \cite{Pujolas:2011he}. This is done by projecting the stress–energy tensor, Eq.\ \eqref{scalarstress}, with respect to the four–velocity $u^\mu$ of comoving observers. The effective energy density and the effective pressure are then defined as 
\begin{align}
\rho_\phi &= T^{(\phi)}_{\mu\nu} u^\mu u^\nu \label{eq:effective-density-KGB} \, , \\
P_\phi &= \tfrac{1}{3} T^{(\phi)}_{\mu\nu} h^{\mu\nu}  \, , 
    \label{eq:effective-pressure-KGB} 
\end{align}
with $h^{\mu\nu} = g^{\mu\nu} + u^\mu u^\nu$ being the spatial projection tensor.

 It has been shown in \cite{Bellini:2014fua} that the complete dynamics of linear cosmological perturbations in scalar-tensor theories belonging to the Horndeski class of actions can be characterized by the background expansion history (i.e.\ $\mathcal{H}$, or $\rho_\phi$, or $w_\phi\equiv\rho_\phi/P_\phi$), the present-day densities, and four independent, time-dependent functions, namely $\alpha_{\rm K}$ (\textit{kineticity}),  $\alpha_{\rm B}$ (\textit{braiding}), $\alpha_{\rm M}$ (\textit{Planck-mass run rate}) and $\alpha_{\rm T}$ (\textit{tensor speed excess}). In the case of $k$-essence and KGB  models, the latter two functions vanish, leaving only $\alpha_{\rm K}$ and $\alpha_{\rm B}$ being defined as
\begin{align}
    \mathcal{H}^2M_{\rm Pl}^2 \alpha_{\rm K} &=  2\bar{X}a^2\Big(  G_{2X}  + 2\bar{X} G_{2XX} -2 G_{3\phi} -2\bar{X}G_{3\phi X}\Big)\nonumber \label{eq:alphak}\\&
    ~\quad+ 12 \bar{\phi}^{'}\bar{X}\mathcal{H}\Big(G_{3X} + \bar{X}G_{3XX}\Big) \, , \\
    \mathcal{H}M_{\rm Pl}^2 \alpha_{\rm B} &= 2\bar{\phi}^{'}\bar{X}G_{3X} \, . \label{eq:alphaB}
\end{align}
where the overbar denotes the background quantities and all the derivatives of the $G$ functions are evaluated at the level of the background in Eqs. \eqref{eq:alphak} and \eqref{eq:alphaB}.

We need to calculate 
 the zeroth-order (background) and first-order perturbations in both the Einstein field equations and the scalar field's equation of motion, as given in Eqs. \eqref{ModifiedEinstein} and \eqref{scalarEOM}. To eliminate the explicit dependence on the $G_2$ and $G_3$ functions, we incorporate the background expressions of  effective dark energy density and pressure, $\bar{\rho}_\phi$ and $\bar{P}_\phi$,  as well as the model parameters $\alpha_{\rm B}$ and $\alpha_{\rm K}$ using  Eqs.\ \eqref{eq:effective-density-KGB} to \eqref{eq:alphaB}. Furthermore, to maintain consistency with \kev, we adopt the following definitions in both notation and implementation
\begin{equation}
\pi = \frac{\delta\phi}{\bar{\phi}^{\prime}} \, ,~\qquad\zeta= \pi^{\prime} + \mathcal{H}\pi -\Psi \, .
\label{eq:zeta_pi}
\end{equation}

\subsubsection{Background level (zeroth-order)}
At the background level, the cosmological dynamics are governed by the modified Friedmann and acceleration equations, with the dark energy field satisfying its own continuity relation. 
Calculating the $^0{}_{0}$ component of the Eq. \eqref{ModifiedEinstein} leads to the modified Friedmann equation
\begin{equation}
\frac{3 M_\text{Pl}^2 \mathcal{H} ^2}{a^2} =\bar{\rho} _{m}{} + \bar{\rho}_\phi   \, ,
\label{Friedmann}
\end{equation}
and the trace of the $^i{}_{j}$ component yields the acceleration equation
\begin{equation}
\frac{M_\text{Pl}^2 \mathcal{H} ^2}{a^2} + \frac{2 M_\text{Pl}^2 \mathcal{H} ^{\prime}}{a^2} =-( \bar{P}_{m}{} + \bar{P}_\phi) \, .
\label{acceleration}
\end{equation}
where $\bar{\rho}_\phi$ and $\bar{P}_\phi$ are the background (zeroth-order) scalar field energy density and pressure, obtained from Eqs. \eqref{eq:effective-pressure-KGB} and \eqref{eq:effective-density-KGB} after restricting to the homogeneous limit of the field
\begin{align}
 \bar{\rho}_\phi &=  -G_2 + \frac{\bar{\phi}{'}^2{G_2}_{X}}{a^2}  - \frac{\bar{\phi}{'}^2{G_3}_{\phi}}{a^2} + \frac{3 \bar{\phi}{{'}}^3\mathcal{H} {G_3}_{X}}{a^4} \, , \label{eq:back-effective-density-KGB} \\
 \bar{P}_\phi &= G_2-\frac{\bar{\phi}{'}^2 {G_3}_{\phi}}{a^2}-\frac{2\bar{\phi}{'}^2G_{3X}(\bar{\phi}^{''}-\mathcal{H}\bar{\phi}^{'})}{a^4} \, .
    \label{eq:back-effective-pressure-KGB} 
\end{align}

The dark energy scalar field equation at the background level takes the form of the continuity equation
\begin{equation}
\bar{\rho}_{\phi}^{\prime} + 3  \mathcal{H} (\bar{\rho}_\phi  + 3\bar{P}_\phi)  = 0 \, ,
\label{eq:contiDE}
\end{equation}
As can be seen explicitly in the  expressions in the next section, the coefficients of the linear perturbation equations can be expressed in terms of $\bar{\rho}_\phi$, $\bar{P}_\phi$ and the $\alpha$ functions, without requiring explicit knowledge of the background evolution of $\phi$. In practice,
\hiclass fully takes care of the background sector. It has two main options: (i) either to solve for the proper evolution of the scalar field using Eq.~\eqref{eq:contiDE} together with Eqs.~(\ref{eq:back-effective-density-KGB}-\ref{eq:back-effective-pressure-KGB}) for a specified theory in which the $G_i$ functions are explicitly defined, or (ii) to parametrize the expansion history directly, which is the spirit of the EFT of dark energy. Here, we choose the latter option and fix the background evolution using the Chevallier-Polarski-Linder (CPL) \citep{Chevallier:2000qy, Linder:2002et} parametrization of the equation of state, i.e.\ $w(a)=w_0+w_a(1-a)$, together with the present-day dark energy fraction $\Omega_{\rm DE}$, which determines $\bar{\rho}_{\phi}$ at $a=1$.
\subsubsection{Linear perturbation (first-order)}
In the derivation of the first‐order perturbation equations we subtract the background Einstein and scalar field equations from the full equations and retain only terms linear in the perturbations. Doing so, the Hamiltonian constraint (Einstein $^0{}_{0}$ component) takes the form
\begin{equation}
    \frac{2M_{\rm Pl}^2}{a^2}\big(\nabla^2\Phi -3\mathcal{H}^2 \Psi -3 \mathcal{H}\Phi^{'}  \big) = \delta\rho_{\rm m} + \delta\rho_{\phi}  \, ,
\end{equation}
where 
\begin{equation}
\begin{aligned}
\delta\rho_{\phi}  &=   - \frac{M_\textrm{Pl}{}^2}{a^2} \bigg\{\alpha_\textrm{B} \mathcal{H}  (\nabla ^{2}{}\pi)  + 3 \alpha_\textrm{B} \mathcal{H} ^2 \Psi  - (3 \alpha_\textrm{B}  + \alpha_\textrm{K}) \mathcal{H} ^2 \zeta  + 3 \alpha_\textrm{B} \mathcal{H}  \Phi ^{\prime}
    \\&\hspace{2.2cm}+\Big[ \alpha_\textrm{B} \mathcal{H} ^{\prime}-  \alpha_\textrm{B} \mathcal{H} ^2   +\frac{a^2}{M_\textrm{Pl}{}^2} (\bar{\rho}_\phi  + \bar{P}_\phi )\Big]3 \mathcal{H} \pi \bigg\} \, .
    \label{eq:deltarhoDE}
\end{aligned}
\end{equation}
\begin{table}[htbp]
\centering
\hspace{-0.5cm}
\caption{Linear equations for stress-energy tensor and scalar field equation of motion in KGB model.}
\label{tab:kgb-equations}
\begin{tcolorbox}[colframe=black, colback=white!10, boxrule=0.5mm, sharp corners=south, title=\textbf{KGB equations}, halign title=center] \textbf{Stress-energy tensor: } 
\begin{equation} \begin{aligned}
T^0_0 &= - \bar{\rho}_\phi + \frac{M_\textrm{Pl}{}^2}{a^2} \bigg\{\alpha_\textrm{B} \mathcal{H}(\nabla^{2}\pi) + 3 \alpha_\textrm{B} \mathcal{H}^2 \Psi - (3 \alpha_\textrm{B} + \alpha_\textrm{K}) \mathcal{H}^2 \zeta + 3 \alpha_\textrm{B} \mathcal{H} \Phi^{\prime} \\ &\hspace{3.cm}+\Big[ \alpha_\textrm{B} \mathcal{H}^{\prime}- \alpha_\textrm{B} \mathcal{H}^2 +\frac{a^2}{M_\textrm{Pl}{}^2} (\bar{\rho}_\phi + \bar{P}_\phi )\Big]3 \mathcal{H} \pi \bigg\} \, , \\[8pt] T^0_i &=-(\bar{\rho}_\phi+\bar{P}_\phi)\partial_{i}\pi +\frac{M_\textrm{Pl}{}^2 \alpha_\textrm{B}\mathcal{H}}{a^2}\partial_i \zeta  \, , \\[8pt] T^i_j &= \big(\bar{P}_\phi + \pi \bar{P}_{\phi}^{\prime}\big) \delta^i_j  \\ &\quad - \frac{M_\textrm{Pl}{}^2}{a^2}\left\{\alpha_\textrm{B} \mathcal{H}\zeta^{\prime} +\zeta \left[2 \alpha_\textrm{B} \mathcal{H}^2 + \mathcal{H}\alpha_\textrm{B}^{\prime} + \alpha_\textrm{B}\mathcal{H}^{\prime} - \frac{a^2}{M_\textrm{Pl}^2}(\bar{\rho}_\phi + \bar{P}_\phi)\right]\right\}\delta^i_j  \, . \end{aligned}\label{eq:DE-Tmunu} \end{equation} \textbf{Equation of motion:} \begin{equation} \textcolor{BrickRed}{A_{\nabla^2 \Psi}} \nabla^2 \Psi + \textcolor{BrickRed}{A_{\zeta'}} \zeta' + \textcolor{BrickRed}{A_{\delta P_\textrm{m}}} \delta P_\textrm{m} + \textcolor{BrickRed}{A_{\Phi'}} \Phi' + \textcolor{BrickRed}{A_{\nabla^2 \pi}} \nabla^2 \pi + \textcolor{BrickRed}{A_{\Psi}} \Psi + \textcolor{BrickRed}{A_{\pi}} \pi + \textcolor{BrickRed}{A_{\zeta}} \zeta = 0 \, , \label{eq:linearEOM} \end{equation} where \begin{align*} \textcolor{BrickRed}{A_{\nabla^2 \Psi}} &= -\frac{\alpha_\textrm{B}}{\mathcal{H}} ~,~\textcolor{BrickRed}{A_{\zeta'}} = \frac{3}{2} \alpha_\textrm{B}^2 + \alpha_\textrm{K}~,~ \textcolor{BrickRed}{A_{\delta P_\textrm{m}}} = -\frac{3 \alpha_\textrm{B} a^2}{2 M_{\text{Pl}}^2 \mathcal{H}}\, ,\\ \textcolor{BrickRed}{A_{\Phi'}} &= -\frac{3 \alpha_\textrm{B}'}{\mathcal{H}} + \alpha_\textrm{B} \left(3 - \frac{3 \mathcal{H}'}{\mathcal{H}^2}\right) + \frac{a^2}{M^2_{\rm Pl}} \left(-\frac{3 \bar{\rho}_\phi}{\mathcal{H}^2} - \frac{3 \bar{P}_\phi}{\mathcal{H}^2}\right) \, , \\ \textcolor{BrickRed}{A_{\nabla^2 \pi}}& = -\frac{\alpha_\textrm{B}'}{\mathcal{H}} - \frac{\alpha_\textrm{B} \mathcal{H}'}{\mathcal{H}^2} + \frac{a^2}{M^2_{\rm Pl}} \left(-\frac{\bar{\rho}_\phi}{\mathcal{H}^2} - \frac{\bar{P}_\phi}{\mathcal{H}^2}\right)\, ,\\ \textcolor{BrickRed}{A_{\Psi}} &= -3 \alpha_\textrm{B}' + \alpha_\textrm{B} \left(3 \mathcal{H} - \frac{3 \mathcal{H}'}{\mathcal{H}}\right) + \frac{a^2}{M^2_{\rm Pl}} \left(-\frac{3 \bar{\rho}_\phi}{\mathcal{H}} - \frac{3 \bar{P}_\phi}{\mathcal{H}}\right)\, ,\\ \textcolor{BrickRed}{A_{\pi}}&= \alpha_\textrm{B}' \left(3 \mathcal{H} - \frac{3 \mathcal{H}'}{\mathcal{H}}\right) + \alpha_\textrm{B} \left(-\frac{3 \mathcal{H}''}{\mathcal{H}} + 9 \mathcal{H}' - \frac{3 \mathcal{H}'^2}{\mathcal{H}^2} - \frac{3 a^2 \bar{P}'_\phi}{2M^2_{\rm Pl} \mathcal{H}}\right) \, , \\ &~\quad+ \frac{a^2}{M^2_{\rm Pl}} \left(3 \bar{\rho}_\phi - \frac{3 \mathcal{H}' \bar{\rho}_\phi}{\mathcal{H}^2} + 3 \bar{P}_\phi - \frac{3 \mathcal{H}' \bar{P}_\phi}{\mathcal{H}^2}\right)\, ,\\ \textcolor{BrickRed}{A_{\zeta}} &= \alpha_\textrm{K}' + \alpha_\textrm{B}^2 \left(3 \mathcal{H} + \frac{3 \mathcal{H}'}{2 \mathcal{H}}\right) + \alpha_\textrm{K} \left(\mathcal{H} + \frac{2 \mathcal{H}'}{\mathcal{H}}\right) + \alpha_\textrm{B} \left[\frac{3}{2} \alpha_\textrm{B}' + \frac{a^2}{M^2_{\rm Pl}} \left(-\frac{3 \bar{\rho}_\phi}{2 \mathcal{H}} - \frac{3 \bar{P}_\phi}{2 \mathcal{H}}\right)\right]\, . \end{align*} 
\end{tcolorbox}
\end{table}
The momentum constraint (Einstein $^0{}_{i}$ component) reads as
\begin{equation}
-\frac{2M_\textrm{Pl}{}^2 }{a^2}\Big( \frac{1}{4}\nabla^2 B_{i} + \partial_{i}\Phi'
+ \mathcal{H}\partial_{i}\Psi
\Big) = \delta T_{i}^{0 (\rm m)} + \delta T_{i}^{0 (\phi)} \, ,
\end{equation}
where
\begin{equation}
 \delta T_{i}^{0 (\phi)} = -(\bar{\rho}_\phi+\bar{P}_\phi)\partial _{i}\pi +\Big(\frac{M_\textrm{Pl}{}^2 \alpha_\textrm{B}\mathcal{H}}{a^2}\Big)\partial_i \zeta \, .
\end{equation}

The pressure equation (spatial trace part of the Einstein equations) takes the form
\begin{equation}
    \frac{2 M_{\rm Pl}^2}{a^2}\big(\Phi^{''}+2\mathcal{H}\Phi^{'} + \mathcal{H}\Psi^{\prime} +\mathcal{H}^2\Psi + 2\mathcal{H}^{'}\Psi  \big) = \delta P_{\rm m} + \delta P_{\phi} \, ,
    \label{eq:spatialTrace}
\end{equation}
where
\begin{equation}
    \delta P_{\phi} =  \pi  P_{\phi}^{\prime} - \frac{M_\textrm{Pl}{}^2}{a^2}\left\{\alpha_\textrm{B} \mathcal{H}  \zeta ^{\prime} +\zeta \bigg[2 \alpha_\textrm{B} \mathcal{H} ^2 +   \mathcal{H}  \alpha_\textrm{B}^{\prime}  +  \alpha_\textrm{B} \mathcal{H}^{\prime} - \frac{a^2}{M_\textrm{Pl}^2}(\bar{\rho}_\phi  + \bar{P}_\phi)\bigg]\right\} \, .
\end{equation}

In summary, the scalar field’s stress-energy tensor and equation of motion are given in Table \ref{tab:kgb-equations}. Note that in deriving Eq. \eqref{eq:linearEOM}, for the sake of implementation, we have removed the dependency on the second time derivative of the metric perturbation $\Phi^{\prime\prime}$ using the  spatial trace part of Einstein equation \eqref{eq:spatialTrace}, which is why the term ${A_{\delta P_\textrm{m}}} \delta P_\textrm{m}$ appears in the scalar field equation. In Appendix \ref{app:k-esselim} we show Eqs. \eqref{eq:DE-Tmunu} and \eqref{eq:linearEOM} reduce to the $k$-essence case when $\alpha_{\rm B}=0$, and we validate our implementation in this limit against the \kev code.

\section{Implementation in \gev}
\label{sec:implementation}
\subsection{Background treatment}
In this first version of the code, \kgb should be compiled with \hiclass in order to retrieve all the background quantities, such as $H(a)$, $\bar\rho_\phi(a)$, $\bar P_\phi(a)$, and the time‑dependent $\alpha_{\rm K}$ and $\alpha_{\rm B}$ functions. \kgb then reads these tables at runtime, converts the physical Hubble into conformal time and transform it into the code units, and advances the scale factor $a(\tau)$ by integrating 

\begin{equation}
 \frac{da}{d\tau} = a\,\mathcal H(a) \, ,  
\end{equation}
using a fourth‑order Runge–Kutta scheme.
\subsection{Perturbation treatment}

In \gev, the Einstein field equations are expanded up to relevant quadratic order in the weak field approximation \cite{Adamek:2016zes}. As a result, they read
\begin{equation}
(1+2 \Phi) \nabla^2 \Phi-3 \mathcal{H} \Phi^{\prime}-3 \mathcal{H}^2(\Phi-\chi)-\frac{1}{2} \delta^{i j} \partial_i \Phi \partial_j \Phi=-4 \pi G a^2 \delta T_0^0 \, ,
\label{eq:gev-Poiss}
\end{equation}
\begin{equation}
    \nabla^4 B_i = 16 \pi G a^2 P^j_i T^0_j\, ,
\end{equation}
\begin{equation}
  \nabla^4 \chi-\left(3 \delta^{i k} \delta^{j l} \frac{\partial^2}{\partial x^k \partial x^l}-\delta^{i j} \nabla^2\right) \Phi_{, i} \Phi_{, j}=4 \pi G a^2\left(3 \delta^{i k} \frac{\partial^2}{\partial x^j \partial x^k}-\delta_j^i \nabla^2\right) T_i^j  \, ,
  \label{eq:gev-Traceless-scalar}
\end{equation}
\begin{equation}
\nabla^4\left(h_{i j}^{\prime \prime}+2 \mathcal{H} h_{i j}^{\prime}-\nabla^2 h_{i j}\right)-4\left(P_i^k P_j^l-\frac{1}{2} P_{i j} P^{k l}\right) \Phi_{, k} \Phi_{, l}=16 \pi G a^2\left(P_{i k} P_j^l-\frac{1}{2} P_{i j} P_k^l\right) T_l^k\, ,
\label{eq:gev-TT-tensor}
\end{equation}
where $\chi = \Phi-\Psi$, and $P_{ij}$ is the transverse projection operator defined as 
\begin{equation}
    P_{ij} = \frac{\partial^2}{\partial x^i \partial x^j} - \delta_{ij}\nabla^2.
\end{equation}
In vanilla \gev, the stress-energy tensor $T^\mu_\nu$ includes only contributions from matter species. As a first step in incorporating the KGB dark energy models into this framework, we consider only the linearized form of the dark energy stress-energy tensor as described in Eq. \eqref{eq:DE-Tmunu}. This linear contribution is then added to the total stress-energy tensor defined in the code.\footnote{Note that in the \gev scheme,  solving the $^0{}_{0}$  constraint for $\Phi$ together with the traceless  $^i{}_{j}$ equations already fixes $\Phi$ and the relativistic quantities $\chi,B_i,h_{ij}$. The remaining four Einstein equations -- the three $^0{}_{i}$ relations and the spatial trace -- are not independent; they follow from those two plus covariant conservation $\nabla_\mu T^{\mu\nu}=0$. Accordingly, the $^0{}_{i}$ equations are used as a consistency check rather than as independent constraints. This redundancy is unchanged when adding our linear KGB sector. We therefore continue to solve only Eqs.\ \eqref{eq:gev-Poiss}–\eqref{eq:gev-Traceless-scalar}-\eqref{eq:gev-TT-tensor} and do not implement the $^0{}_{i}$ part of the KGB stress-energy tensor in this first version of the code.} The same is true for the equation of motion of the scalar field, where we only consider the linear description for the evolution of $\pi$, given by Eq. \eqref{eq:linearEOM}. We restrict ourselves to the linear case in this work for two reasons. On the one hand, the nonlinear regime is significantly more complex and involves theoretical and numerical difficulties. In related studies such as the implementation of the $k$-essence model in \gev, it has been shown that certain nonlinear terms can introduce instabilities \cite{Hassani:2021tdd}. On the other hand, that work also showed that for sufficiently large speed of sound the linear terms in the scalar field equations were able to capture the dynamics correctly so that we expect our results to be relevant in part of the parameter space where braiding is sufficiently small and the speed of sound is high. We will investigate the nonlinear self-interactions of dark energy in detail in a dedicated follow up study.

The KGB scalar field dark energy sources the Poisson equation \eqref{eq:gev-Poiss} via $\delta\rho_\phi$ and the spatial trace equation
via $\delta P _\phi$.
At the linear level, $T^i_j$ is proportional to $\delta^i_j$, see \eqref{eq:DE-Tmunu}. Consequently, it gives no direct source to the traceless scalar equation \eqref{eq:gev-Traceless-scalar} or transverse-traceless tensor equation \eqref{eq:gev-TT-tensor} in which the right hand side projectors remove the isotropic stresses. Nevertheless, the modified $\Phi$ and $\Psi$ obtained from the Poisson equation feed back on the left hand sides of the traceless scalar and transverse-traceless equations so these sectors are indirectly affected through the metric potentials.

It is important to mention that, although the nonlinear terms in the scalar field equation are neglected, the energy density of the scalar field still exhibits nonlinear behavior, as it is sourced by the fully nonlinear matter.

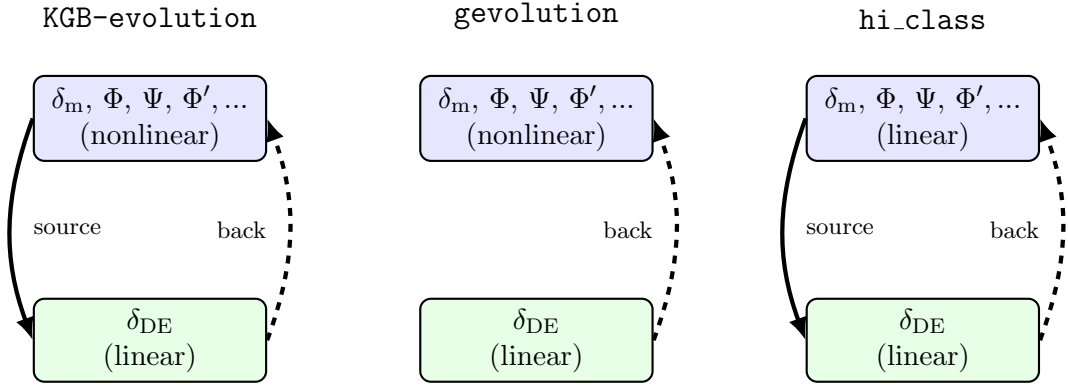
\begin{figure}[htbp]
  \centering
  \begin{tikzpicture}[
    block/.style={
      draw,
      thick,
      rounded corners,
      text width=2.8cm,
      align=center,
      minimum height=1.1cm
    },
    arrow source/.style={
      ultra thick,
      bend right=20,
      -{Latex[length=3mm,width=2.5mm]}
    },
    arrow back/.style={
      ultra thick,
      dashed,
      bend right=20,
      -{Latex[length=3mm,width=2.5mm]}
    }
  ]
    \matrix (m) [matrix of nodes,
      column sep=2cm,
      row sep=1.8cm,
      row 1/.style={nodes={block, fill=blue!10}},
      row 2/.style={nodes={block, fill=green!10}}
    ]{
      {\(\delta_{\rm m},\,\Phi,\, \Psi,\,\Phi',...\)\\(nonlinear)}
      &
      {\(\delta_{\rm m},\,\Phi,\, \Psi,\,\Phi',...\)\\(nonlinear)}
      &
      {\(\delta_{\rm m},\,\Phi,\,\Psi,\,\Phi',...\)\\(linear)}  \\
      {\(\delta_{\rm DE}\)\\(linear)}
      &
      {\(\delta_{\rm DE}\)\\(linear)}
      &
      {\(\delta_{\rm DE}\)\\(linear)}  \\
    };

    \node[font=\large\bfseries, above=5mm of m-1-1] {\kgb};
    \node[font=\large\bfseries, above=4.7mm of m-1-2] {\gev};
    \node[font=\large\bfseries, above=4.7mm of m-1-3] {\hiclass};

    \draw[arrow source]
      (m-1-1.west) 
      to node[right=1.5mm,font=\footnotesize]{source} (m-2-1.west);
    \draw[arrow back]
      (m-2-1.east) 
      to node[left=1.5mm,font=\footnotesize]{back}   (m-1-1.east);

    \draw[arrow back]
      (m-2-2.east) 
      to node[left=1.5mm,font=\footnotesize]{back}   (m-1-2.east);

    \draw[arrow source]
      (m-1-3.west) 
      to node[right=1.5mm,font=\footnotesize]{source} (m-2-3.west);
    \draw[arrow back]
      (m-2-3.east) 
      to node[left=1.5mm,font=\footnotesize]{back}   (m-1-3.east);

  \end{tikzpicture}
  \caption{%
    (a) \kgb: nonlinear metric perturbations directly (via metric potentials and their time derivatives) source a linear dark energy perturbation and vice versa. 
    (b) \gev: purely linear $\delta_{\rm DE}$ sources metric perturbations.
    (c) \hiclass: all perturbations are treated linearly both ways.
  }
  \label{fig:three-evolutions}
\end{figure}

In Fig. \ref{fig:three-evolutions} we show schematically how different components in different codes are treated and interact with each other. In the \kgb scheme, the nonlinear matter perturbation sources 
all the metric perturbations. These nonlinear fields then source the dark energy perturbation $\delta_{\rm DE}$, which is formally kept at linear order but acquires effective nonlinear corrections through this coupling. The resulting dark energy density perturbations, in turn, feed back into the evolution of the gravitational potentials, together with the matter perturbations. The same scheme is applied in \hiclass, where however all the components are kept at linear order. In the \gev scheme, dark energy perturbations can be included as a linear realization\footnote{In \gev, the fluid approach of \class enables a direct linear realization of the scalar field energy density for $k$‑essence models, see \citep{Hassani:2019lmy}; by contrast, creating a realization of the KGB energy density is more involved since \class lacks a fluid description of KGB and \hiclass does not directly output $\delta_{\rm DE}$. Consequently, one has to manually construct the linear realization of the KGB energy density in \gev, see Appendix \ref{sec:deltaDE} for more details.} from \hiclass and thus it is not sourced by the nonlinear matter and metric perturbations (rather, it is sourced by linear matter and metric perturbations through \hiclass) and provides a linear source term for those nonlinear perturbations.

\subsection{Options for dark energy models}

To enable a flexible configuration of the dark energy sector, we have added the following configuration options which are similar to those of the \hiclass code. Table \ref{tab:de-kgb-options} shows all the new dark energy and KGB settings that can be adjusted in the \texttt{settings.ini} file within \kgb. In brief, the \texttt{gravity\_model} option determines the time‐dependence of the EFT functions $\alpha_i$; choosing \texttt{propto\_omega}, makes the $\alpha_i$ functions to grow proportionally to the dark energy density: $\alpha_i=\hat\alpha_i\,\Omega_{\rm DE}(\tau)$, \texttt{propto\_scale} scales the evolution of $\alpha_i$ with the scale factor: $\alpha_i=\hat\alpha_i\,a(\tau)$, and \texttt{constant\_alphas} keeps $\alpha_i=\hat\alpha_i$ at a constant value. It is important to mention that, in principle, all parametrizations of the $\alpha_i$ functions are phenomenological, rather than uniquely fixed by a fundamental theory.
The \texttt{parameters\_smg} entry includes the amplitudes $(\hat\alpha_{\rm K},\hat\alpha_{\rm B},\hat\alpha_{\rm M},\hat\alpha_{\rm T})$ and the initial Planck mass squared $M^2_{*,\rm ini}$. Note that in our case, the KGB model, $\hat\alpha_{\rm M}$ and $\hat\alpha_{\rm T}$ are set to zero by default and  $M^2_{*,\rm ini}$ is fixed to 1. The background expansion is determined by \texttt{expansion\_model}; currently \texttt{wowa} is supported with the corresponding CPL equation of state parameters $(w_0,w_a)$ given by \texttt{expansion\_smg}. More generally, any time-dependence for $\alpha_i(\tau)$ and any expansion history supported by \hiclass can be used. If a different EFT model is desired, one can implement it on the \hiclass side; \kgb{} will then  read and use the corresponding $\alpha_i$ and background quantities through \texttt{gravity\_model} and \texttt{expansion\_model} automatically.

On the numerical side, \texttt{n\_kgb\_numsteps} controls how many sub-steps are used to integrate the KGB scalar field within each main time step of the code--for gravity solver and particle update--under the assumption that particles and the gravitational potential remain unchanged during the scalar field sub-step integration. This assumption holds for slowly moving matter species (e.g., dark matter), but may break down for fast-moving particles such as massive neutrinos, in which case the overall time step of simulations should be decreased. 
The time stepper sets the conformal time step $\Delta\tau$ as the minimum of two bounds
\begin{equation}
\Delta\tau = \min\left(\, {\rm{CF}}\times\Delta {\rm x},\frac{\texttt{time\_step\_limit}}{\mathcal H(a)}\right),  
\end{equation}
where $\Delta {\rm x} = \texttt{boxsize}/\texttt{Ngrid}$ is the comoving spatial resolution and $\rm{CF}$ is a fixed Courant factor chosen by the user, which in principle measures the number of grid cell units that a \textit{light} signal traverses during a single time step. Since the velocity of CDM particles is usually less than $1\%$ of the speed of light, choosing $\rm CF \leq 100$ keeps their displacement below one grid cell in each time step.\footnote{For performance monitoring, the code outputs a “CDM Courant factor”, defined as ${\rm{CF}^{Max}_{CDM}} = \mathrm{CF}\times \max(v_{\rm DM})$. This value indicates how far the fastest dark matter particle moves during one integration step relative to the grid unit; if it exceeds unity, some particles may cross multiple cells per step, reducing the accuracy of their orbit integration.} 
The global time step limit overrides the Courant factor condition whenever ${\texttt{time\_step\_limit}}/{\mathcal H(a)}$ is smaller, so that large scale evolution remains accurate.

After $\Delta\tau$ is fixed in the above condition,  we can compute the Courant factor for the dark energy scalar field using the speed of sound for the scalar field ($c_s$)
\begin{equation}
{\rm{CF}_{\phi}} = c_s \times \frac{\Delta\tau}{\Delta {\rm x}}\, .
\end{equation}
Because the scalar field equation of motion is an explicit partial differential equation, its numerical evolution must satisfy a Courant–Friedrichs–Lewy (CFL) stability condition, which requires that no signal propagate farther than about one grid unit during a single integration step, i.e. ${\rm{CF}_{\phi}}\leq 1$.
On the other hand, since $\Delta\tau$ is chosen through CF, to ensure stable evolution of both the gravity solver and the particle dynamics, a scalar field with a propagation speed $c_s$ larger than the typical particle velocities may violate this stability condition. To maintain stability and sufficient temporal resolution for the scalar field evolution, we evolve it using multiple sub-steps within each main time step of the gravity solver. The number of these sub-steps, specified by \texttt{n\_kgb\_numsteps}, is chosen so that the scalar field satisfies the CFL condition.
Therefore, a reasonable choice would be
\begin{equation}
  \texttt{n\_kgb\_numsteps}\ \gtrsim\ \frac{c_s\Delta\tau} {\Delta{\rm x}}\, .
\end{equation}

The parameter \texttt{kgb\_source\_gravity} specifies whether the KGB stress–energy appears in the Poisson equation ($0$=off, $1$=on). 
\begin{table}[htbp]
  \centering
  \caption{Summary of models and numerical parameters}
  \renewcommand{\arraystretch}{0.98}
  \begin{tabularx}{\textwidth}{|>{\centering\arraybackslash}m{4cm}
                               |>{\centering\arraybackslash}m{4cm}
                               |X|}
    \hline
    \rowcolor{gray!30}
    \textbf{Parameter} & \textbf{Value / Option} & \textbf{Description} \\
    \hline\hline
    
    \rowcolor{white}
    \texttt{gravity\_model} & \texttt{propto\_omega} & \(\alpha_i = \hat{\alpha}_i\,\Omega_{\rm DE}(\tau)\) \\
    \cline{2-3}
    \rowcolor{white}
    & \texttt{propto\_scale} & \(\alpha_i = \hat{\alpha}_i\,a(\tau)\) \\
    \cline{2-3}
    \rowcolor{white}
    & \texttt{constant\_alphas} & \(\alpha_i = \hat{\alpha}_i\) \\
    \hline
    
    \rowcolor{gray!10}
    \texttt{parameters\_smg} & \texttt{\((\hat{\alpha}_{\rm K},\hat{\alpha}_{\rm B},\hat{\alpha}_{\rm M},\hat{\alpha}_{\rm T},M^2_{*,\rm ini})\) } &  A vector containing the numerical values of $\alpha_i$ used in the chosen \texttt{gravity\_model} parametrization \\
    \hline
    \rowcolor{white}
    \texttt{expansion\_model} & \texttt{wowa} & Parametrizes the universe’s expansion history via a DE with EoS of $w(a)=w_0 + w_a(1 - a)$\\
    \hline
    \rowcolor{gray!10}
    \texttt{expansion\_smg} & $(w_0, w_a)$ & A vector containing the numerical values of DE EoS parameters. \\
    \hline
    \rowcolor{white}
    \texttt{n\_kgb\_numsteps} & int number  & Number of sub-steps for the KGB scalar evolution. \\
    \hline
    
    \rowcolor{gray!10}
    \texttt{kgb\_source\_gravity} & 0 & DE stress-energy tensor is not included in Poisson equation. \\
    \cline{2-3}
    \rowcolor{gray!10}
    & 1 & DE stress-energy tensor is included in Poisson equation. \\
    \hline
  \end{tabularx}
  \label{tab:de-kgb-options}
\end{table}
 All of these options are summarized in Table \ref{tab:de-kgb-options}. More details about the numerical implementation can be found in Appendix \ref{Numerical Implementation}.

\newpage
\section{Numerical results}
\label{sec:results}
In this section we show the numerical results obtained with \kgb and compare them to \hiclass and \kev codes. Note that in \kgb we can recover the results of $k$-essence (the results of \kev code) by setting $\alpha_{\rm B}$ to zero. All the results for power spectra in this section are obtained using the fixed cosmological parameters given in Table \ref{table:cosmoparams}. To capture both the largest and smallest modes, we run three simulations with different resolutions: (Ngrid, boxsize) = ($3072^3$, 90 Gpc/$h$), ($4096^3$, 15 Gpc/$h$), and ($3072^3$, 4 Gpc/$h$). Since \hiclass is implemented exclusively in synchronous gauge, we employ the transformations in Appendix \ref{sec:GaugeTrans} to express the perturbations in Newtonian gauge for direct comparison with the output of \kgb. Likewise, since \hiclass provides the simulation's initial conditions, we use the same transformations to map them into Newtonian gauge at the initial redshift (here $z=100$).
We fix $w_0=-0.9$, $w_a=0$, and choose $\hat{\alpha}_{\rm K}=3000$ in both the $k$-essence and KGB models so that, by adding $\hat{\alpha}_{\rm B}=1.5$\footnote{Note that the extreme braiding choice $\hat\alpha_{\mathrm B}=1.5$ is ruled out by current observations \citep{Chudaykin:2025gdn}; we use it solely to stress-test \kgb and highlight braiding effects. In the next section we repeat the analysis with more observationally reasonable $\hat\alpha_{\mathrm B}$ values, which yield smaller deviations.} only in the KGB case, we can isolate and study its impact. In both cases we adopt the commonly used \texttt{propto\_omega} parametrization, where each $\alpha$ scales with the dark energy density fraction
\begin{equation}
    \alpha_i(\tau) = \hat{\alpha}_i \Omega_{\rm DE}(\tau) \, .
\end{equation}
All simulations are initialized with identical random seeds, so that stochastic fluctuations due to cosmic variance are correlated across runs. As a result, the cosmic variance is suppressed in relative differences of power spectra from two simulations to reflect only the physical effects under investigation.

\subsection{Power spectra analysis}
\begin{figure}[t]
\centering
\includegraphics[width=1\linewidth]{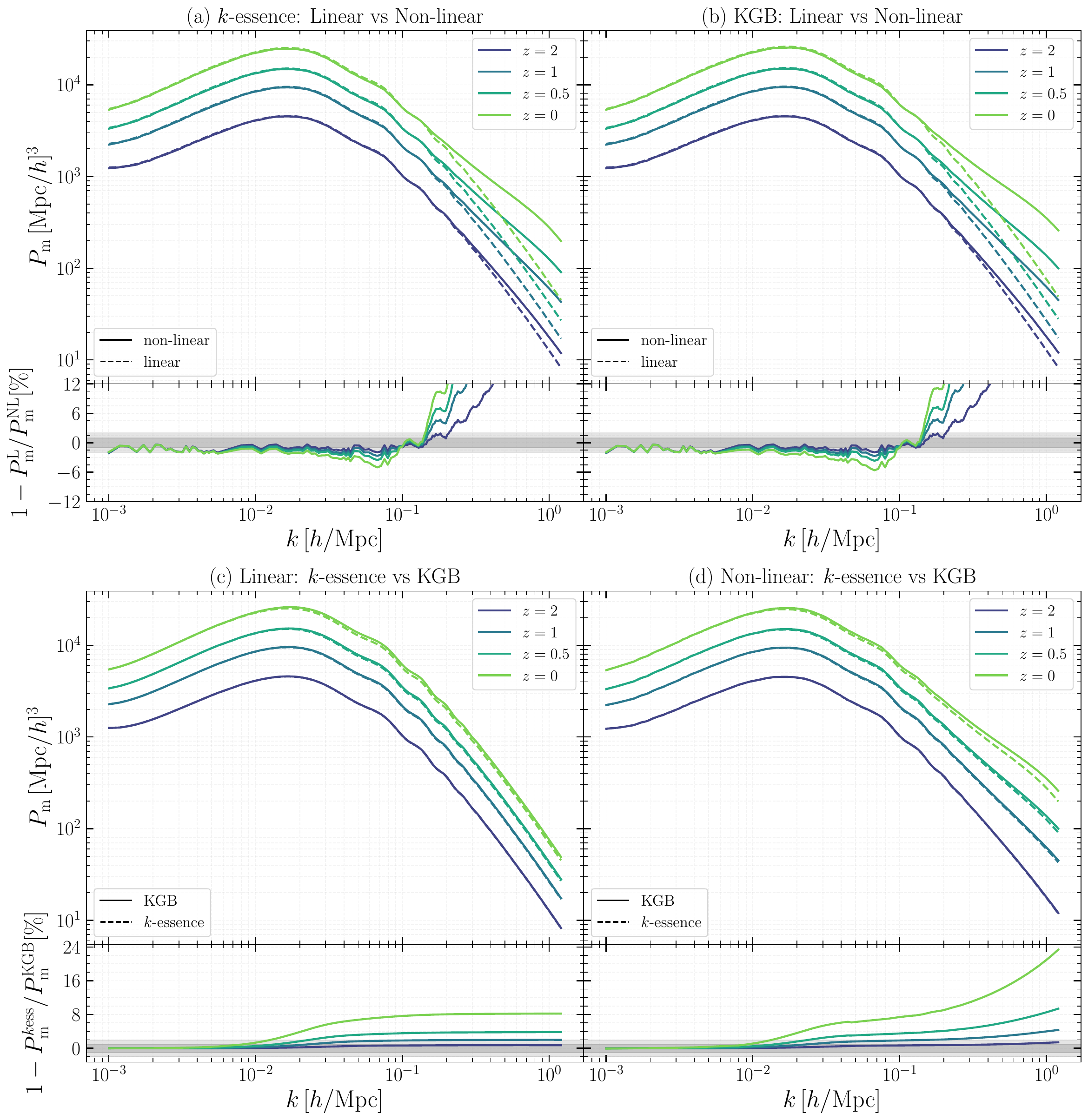}
\caption{Matter power spectra at $z=2, 1, 0.5, 0$, with lower sub-plots showing relative differences.  
(a) $k$-essence model ($\hat\alpha_{\rm K}=3000,\ \hat\alpha_{\rm B}=0$): linear (dashed, \hiclass) vs nonlinear (solid, \kgb).  
(b) KGB model ($\hat\alpha_{\rm K}=3000,\ \hat\alpha_{\rm B}=1.5$): linear (dashed, \hiclass) vs nonlinear (solid, \kgb).  
(c) Linear spectra: $k$-essence (dashed) vs KGB (solid).  
(d) Nonlinear spectra: $k$-essence (dashed) vs KGB (solid). The grey lines indicate 1\% and 2\% bounds.  
}
\label{fig:matter_power_spectrum}
\end{figure}

Fig. \ref{fig:matter_power_spectrum} shows the matter power spectrum at redshifts $z=2$, $1$, $0.5$ and $0$ in four panels.
The upper row, panels (a) and (b), compares the linear (dashed) and nonlinear (solid) spectra within each model, while the lower row, panels (c) and (d), directly compares the two theories in both the linear and nonlinear cases. Each panel includes a small 
lower sub-plot
plotting the relative differences.
From panels (a) and (b) we can see that nonlinear behavior emerges for $k\gtrsim0.1$~Mpc/$h$ at $z= 0$ in both the $k$-essence case ($\hat\alpha_{\mathrm{K}}=3000,\;\hat\alpha_{\mathrm{B}}=0$) and the KGB case ($\hat\alpha_{\mathrm{K}}=3000,\;\hat\alpha_{\mathrm{B}}=1.5$).
At very large scales, the results of simulations agree very well with the linear prediction from \hiclass in both models. This is illustrated in the lower sub-plots where the relative differences between linear and nonlinear predictions remain within 1 percent. On intermediate scales, nonlinear corrections induce a modest suppression of both the KGB and $k$-essence power spectra relative to the linear prediction.  This effect is driven by the negative, dominant one-loop $P_{13}^{\rm SPT}$ contribution in the combination $P_{22}^{\rm SPT}+P_{13}^{\rm SPT}$ at quasi-linear scales (see \cite{Hassani:2019lmy,Jalilvand:2019brk}). Despite the one-loop suppression on intermediate scales, the simulation results begin to depart significantly from linear \hiclass predictions for modes with $k \gtrsim k_{\rm nl}(z)$, where $k_{\rm nl}(z)$ is the redshift-dependent nonlinear scale (e.g., $k_{\rm nl}(0)\sim0.1~h/\mathrm{Mpc}^{-1}$ at $z=0$).

Panels (c) and (d) highlight how the two theories differ in both the linear and nonlinear cases. In the linear \hiclass prediction (panel c), the extreme choice $\hat\alpha_{\mathrm{B}}=1.5$ produces nearly an $4 \%$ deviation from the $k$-essence case at $z=0$ and $k=0.02$ Mpc/$h$. This gap gradually widens toward smaller scales, reaching about $8 \%$ by $k=1$ Mpc/$h$, and as expected, diminishes by going to higher redshifts.
In the fully nonlinear case (panel d), the $\sim$4 \% offset at $z=0$ and $k=0.02$ Mpc/$h$ is preserved but then rises to about 24 \% by $k=1$ Mpc/$h$, highlighting more pronounced nonlinear effects in the KGB model.
\begin{table}[h]
 \caption{Cosmological parameters used for the calculation of the power spectra.}
  \centering
  \begin{tabular}{|*{9}{|c}||}
    \hline
    \rowcolor{crisp}
    $\Omega_{\text{cdm}}$ 
      & $\Omega_\text{DE}$ 
      & $\Omega_\text{b}$ 
      & $\Omega_\text{g}$ 
      & $w_0$ 
      & $w_a$ 
      & $h$ 
      & $A_s$ 
      & $n_s$ \\
    \hline\hline
    0.2638 
      & 0.6878 
      & 0.04828 
      & $5.418\times10^{-5}$ 
      & -0.90 
      & 0 
      & 0.6755 
      & $2.215\times10^{-9}$ 
      & 0.9619 \\
    \hline
  \end{tabular}
  \label{table:cosmoparams}
\end{table}

It is important to mention that we have not employed the \texttt{Halofit} fitting formula in \hiclass for nonlinear corrections, since \texttt{Halofit} has been calibrated exclusively on $\Lambda$CDM simulations and it is not designed to capture the nonlinear behavior of the $k$-essence and KGB models studied here. However, in cases where dark energy clustering is minimal and nonlinear enhancements arise primarily from differences in background evolution, as occurs for some $k$-essence parameters, \texttt{Halofit} can perform reasonably well. This is because it effectively reproduces nonlinearities in the dark matter sector, thereby improving the fit compared to linear predictions.
\begin{figure}[t]
\centering
\includegraphics[width=1\linewidth]{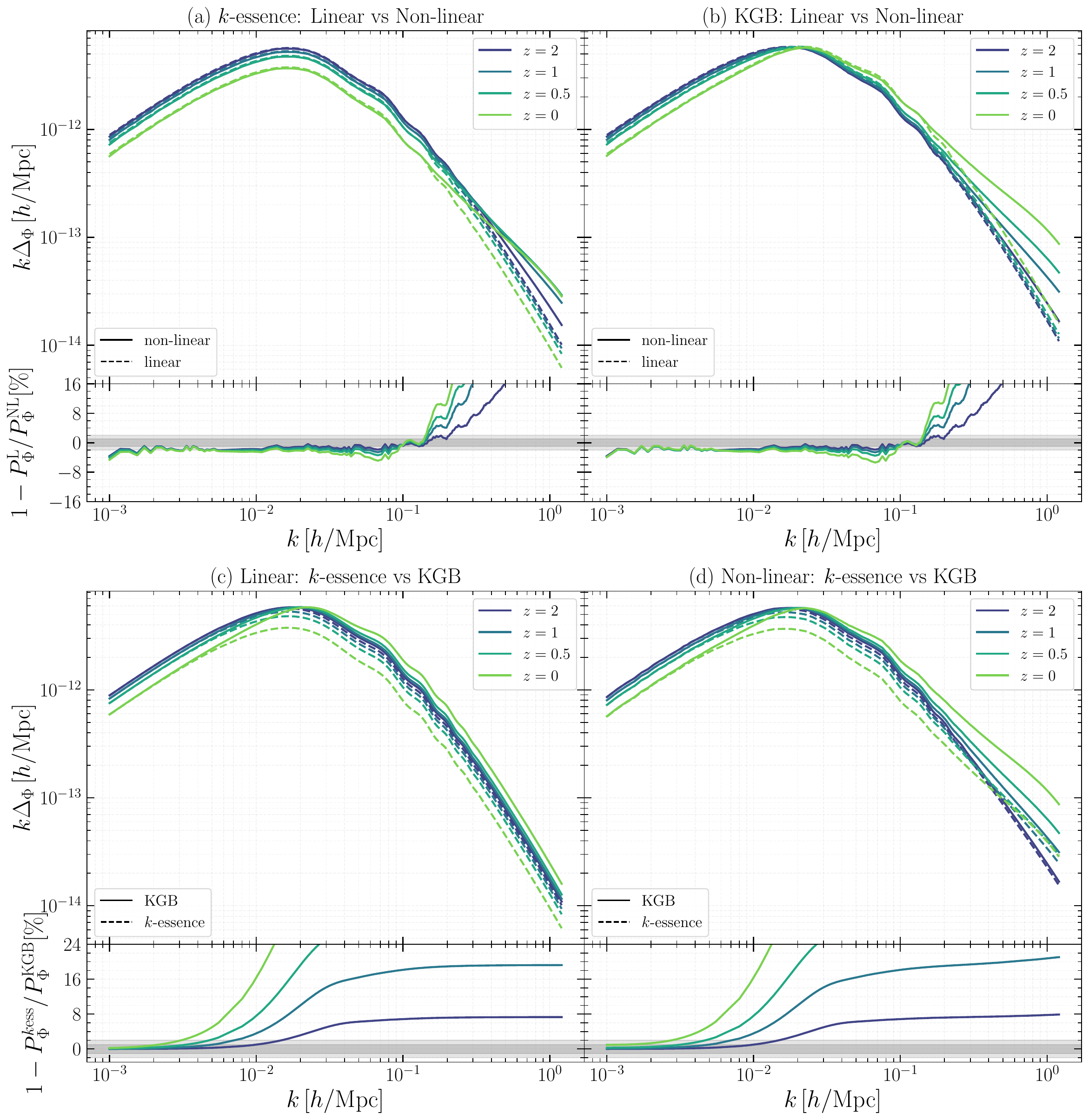}
\caption{
Potential power spectra at $z=2, 1, 0.5, 0$, with lower sub-plots showing relative differences.  
(a) $k$-essence model ($\hat\alpha_{\rm K}=3000,\ \hat\alpha_{\rm B}=0$): linear (dashed, \hiclass) vs nonlinear (solid, \kgb).  
(b) KGB model ($\hat\alpha_{\rm K}=3000,\ \hat\alpha_{\rm B}=1.5$): linear (dashed, \hiclass) vs nonlinear (solid, \kgb).  
(c) Linear spectra: $k$-essence (dashed) vs KGB (solid).  
(d) Nonlinear spectra: $k$-essence (dashed) vs KGB (solid). The grey lines indicate 1\% and 2\% bounds. }
\label{fig:potential_power_spectrum}
\end{figure}

 We have performed the same analysis for the dimensionless potential power spectrum $\Delta_\Phi$, as illustrated in Fig. \ref{fig:potential_power_spectrum}, adopting the general convention
 $$
 \Delta_X(k,z) = \frac{k^3}{2\pi^2}P_X(k,z) \, .
 $$
for any dimensionless field $X$. In the upper panels, at small scales, the results of simulation for both KGB and $k$-essence overshoot the linear prediction from \hiclass due to the nonlinear evolution of matter. In the lower panels, on very large scales, the potential power spectra in both models agree very well, but beyond roughly $ k \simeq  0.004~h/$Mpc, the result of KGB prediction starts to deviate from $k$-essence, predicting higher values for the potential power spectrum.
This overall enhancement of the potential power spectrum in the KGB model arises from the large value of $\hat{\alpha}_{\rm B}$, which produce much larger dark energy perturbations in the KGB model as we will see below. These perturbations source directly the gravitational potential through the Poisson equation or the Hamiltonian constraint. The impact of the dark energy clustering on the matter power spectrum on the other hand is reduced compared to the potential power spectrum, since dark energy does not directly couple to the dark matter. This implies that observables related to the gravitational potential, like gravitational lensing or the Integrated Sachs–Wolfe (ISW) effect, are potentially more powerful in constraining the KGB model, than the matter power spectrum.

\begin{figure}[t]
\centering
\includegraphics[width=1\linewidth]{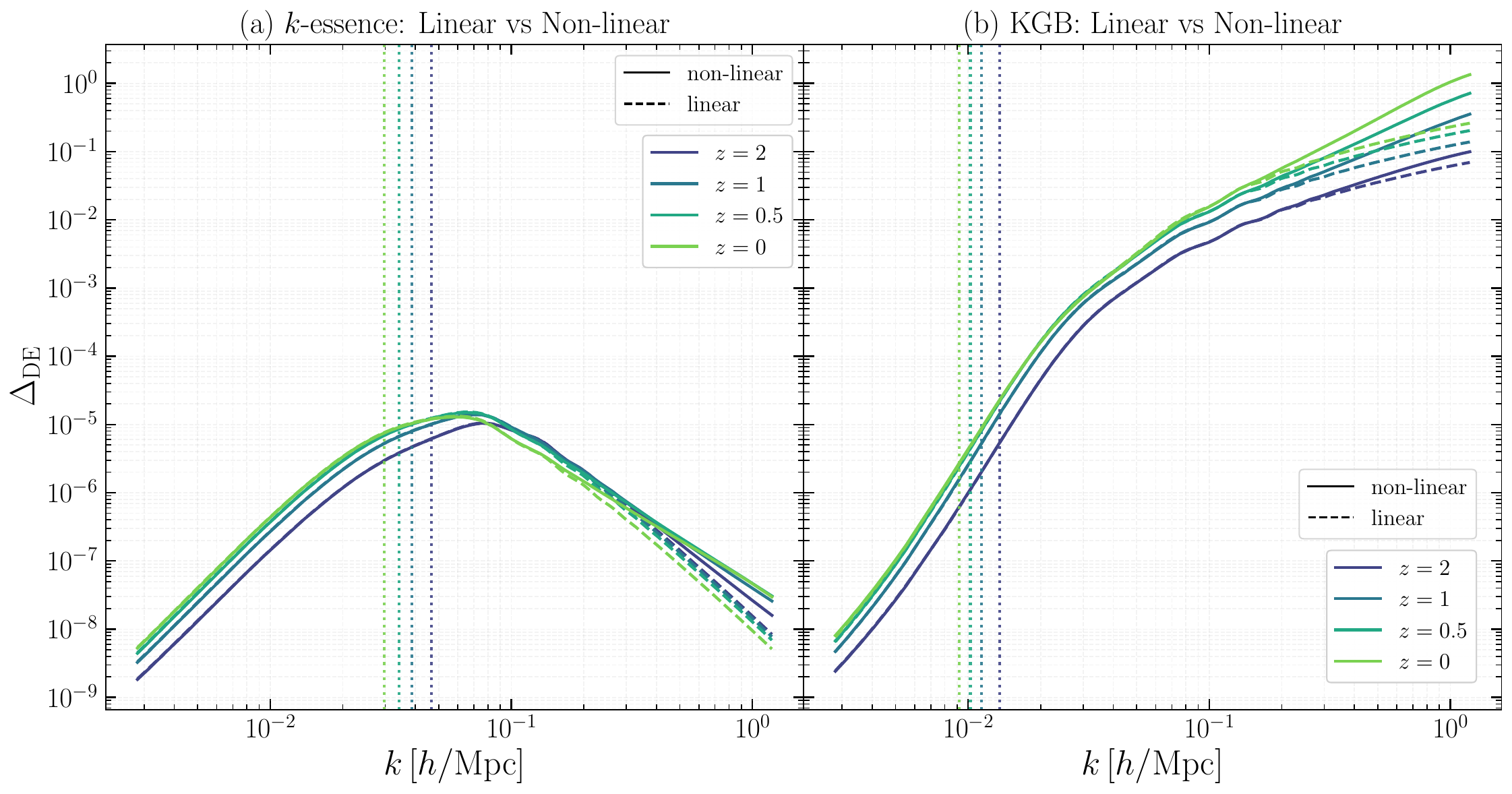}
\caption{Linear (dashed lines) and nonlinear (solid lines) dark energy density contrast power spectra at $z=2,1,0.5, 0$ for (a) $k$-essence model, where modes grow above and decay within the sound horizon (vertical dotted lines), and (b) KGB model, where braiding parameter $\alpha_{\rm B}$, sustains enhanced clustering even inside the horizon.
}
\label{fig:darkenergy_power_spectrum}
\end{figure}

In Fig. \ref{fig:darkenergy_power_spectrum} we compare the dark energy density contrast power spectra (using Eq. \eqref{eq:deltarhoDE})
at different redshifts for both the KGB and $k$-essence models. We remind the reader that we have adopted the \texttt{propto\_omega} parametrization for the $\alpha$ functions, which fixes their time dependence proportional to the dark energy background density. We also show the scale of the sound horizon as vertical dotted lines. We compute the comoving sound horizon at each redshift as
\begin{equation}
 r_s(z)=\int_{0}^{a(z)}
   \frac{c_s(a)}{a^2H(a)}da \, .   
\end{equation}
In the KGB model, the squared speed of sound is given by (see \cite{Bellini:2014fua})
\begin{equation}
    c_s^2(a)
= \frac{-\alpha_{\mathrm{B}}^2\mathcal{H}^2 +2\mathcal{H}\,\alpha'_{\mathrm{B}}
        +2\alpha_{\mathrm{B}}\,\mathcal{H}'
        +\frac{2a^2}{M_{\rm Pl}^2}\bigl(\bar\rho_\phi+\bar P_\phi\bigr)}
{\mathcal{H}^2\bigl(3\alpha_{\mathrm{B}}^2+2\alpha_{\mathrm{K}}\bigr)} \, .
\label{eq:soundspeed}
\end{equation}
Given our fiducial values for cosmological parameters listed in Table \ref{table:cosmoparams}, and the adopted values of $\hat\alpha_{\rm K}=3000,\ $ and $ \hat\alpha_{\rm B}=1.5$, this gives a time varying $c_s^2(a)$ that evolves from $\sim2.9\times10^{-3}c^2$ (where $c$ is the speed of light) at $z=100$ down to $\sim7.7\times10^{-5}c^2$ at $z=0$.
In the $k$-essence limit, $\alpha_{\mathrm{B}}=0$, Eq. \eqref{eq:soundspeed} reduces to
\begin{equation}
 c_s^2
= \frac{a^2\bigl(\bar\rho_\phi+\bar P_\phi\bigr)}
       {M_{\rm Pl}^2\alpha_{\mathrm{K}}\mathcal{H}^2} \, ,   
\end{equation}
which yields a constant throughout cosmic history\footnote{This can be easily shown using $\bar\rho_\phi+\bar P_\phi=M_{\rm Pl}^2(1+w)\,3H^2\Omega_{\rm DE}$, $\alpha_{\rm K}=\hat\alpha_{\rm K}\,\Omega_{\rm DE}$ and $\mathcal{H} = a H$,  so that
$$
c_s^2=\frac{\bar\rho_\phi+\bar P_\phi}{M_{\rm Pl}^2\alpha_{\rm K}H^2}
=\frac{3(1+w)}{\hat\alpha_{\rm K}}\,.
$$}, here $c_s^2 = 10^{-4}c^2$. 

In panel (a) of Fig.\ \ref{fig:darkenergy_power_spectrum}, we can see that for the $k$-essence model, dark energy perturbations grow on scales larger than the sound horizon but begin to decay once they enter it. On smaller scales, nonlinear clustering in the matter component enhances the dark energy power spectrum relative to its linear prediction, which is in excellent agreement with findings in \cite{Hassani:2019lmy}.
By contrast, panel (b) shows that in the KGB model, dark energy perturbations continue to grow rapidly both inside and outside the sound horizon, driven by the braiding term $\alpha_{\mathrm{B}}$. Matter nonlinearities further amplify this growth at small scales, causing the nonlinear power to overshoot the linear \texttt{hi\_class} prediction. For such large values of $\alpha_{\mathrm{B}}$, the dark energy density perturbations in the KGB model are many orders of magnitude larger than those in $k$-essence for scales of the order of the sound horizon or smaller. We will examine this strong clustering behavior in more detail by looking at snapshots for different values of $\hat\alpha_{\rm B}$ in the next subsection.

\begin{figure}[t]
\centering
\includegraphics[width=1\linewidth]{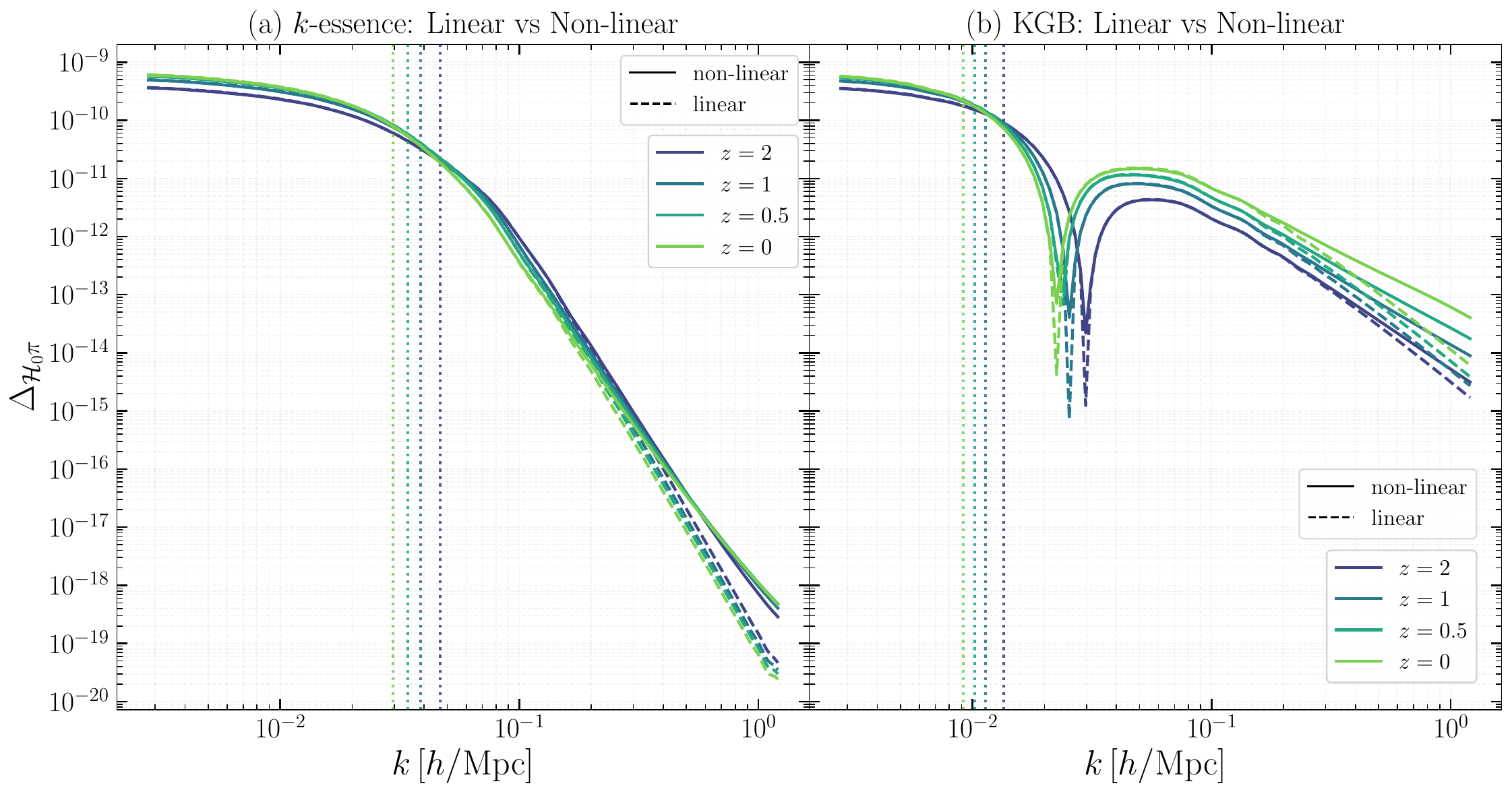}
\caption{Linear (dashed lines) and nonlinear (solid lines) scalar field perturbation power spectra at $z = 2, 1, 0.5, 0$ for (a) $k$-essence and (b) KGB models. In $k$-essence case, the power decays inside the sound horizon and nonlinear contributions exceed linear ones at high $k$ due to $\pi$ being sourced by nonlinear matter clustering. In KGB model, $\pi$ crosses zero at intermediate scales and changes sign, which explains the turnover of the power spectrum at that point. Beyond that point, the spectrum decays gradually toward smaller scales. 
}
\label{fig:pi_power_spectrum}
\end{figure}

In Fig. \ref{fig:pi_power_spectrum} we plot the dimensionless spectrum of the scalar field $\mathcal{H}_0\pi$ perturbation. In both the $k$‑essence and KGB models, the perturbations begin to decay as soon as they enter the sound horizon, but their subsequent behavior differs. In the $k$‑essence case the decay is gradual and continues toward small scales, where matter nonlinearities eventually drive the field into a nonlinear regime, causing the spectrum to overshoot the linear \hiclass prediction. By contrast, in the KGB model, where the sound horizons are shifted to larger scales, the presence of the braiding parameter $\alpha_{\rm B}$  leads to a sharper drop in amplitude inside the horizon, which is due to the zero‑crossing and sign change in the scalar field perturbation. This corresponds to a turnover of the power spectrum at that point, after which, the spectrum continues with a slow decline and  eventually becomes nonlinear at small scales under the effects of nonlinear matter clustering. 

Note that we can recover the velocity divergence $\theta_{\rm DE}(k,z)$ from the scalar field perturbation $\pi$ using $\theta_{\rm DE}(k,z)=k^2\,\pi(k,z)$. Thus, the sign of $\pi(k)$ directly tells us whether the dark energy flow is diverging ($\pi>0$) or converging ($\pi<0$) in comoving space. In the $k$-essence model, $\pi$ stays positive at all scales, so dark energy flow continually diverges. However, in the KGB model, $\pi$ flips from positive to negative, indicating a transition from divergent behavior on large scales to convergent behavior on intermediate and small scales.

\begin{figure}[t]
\centering
\includegraphics[width=1\linewidth]{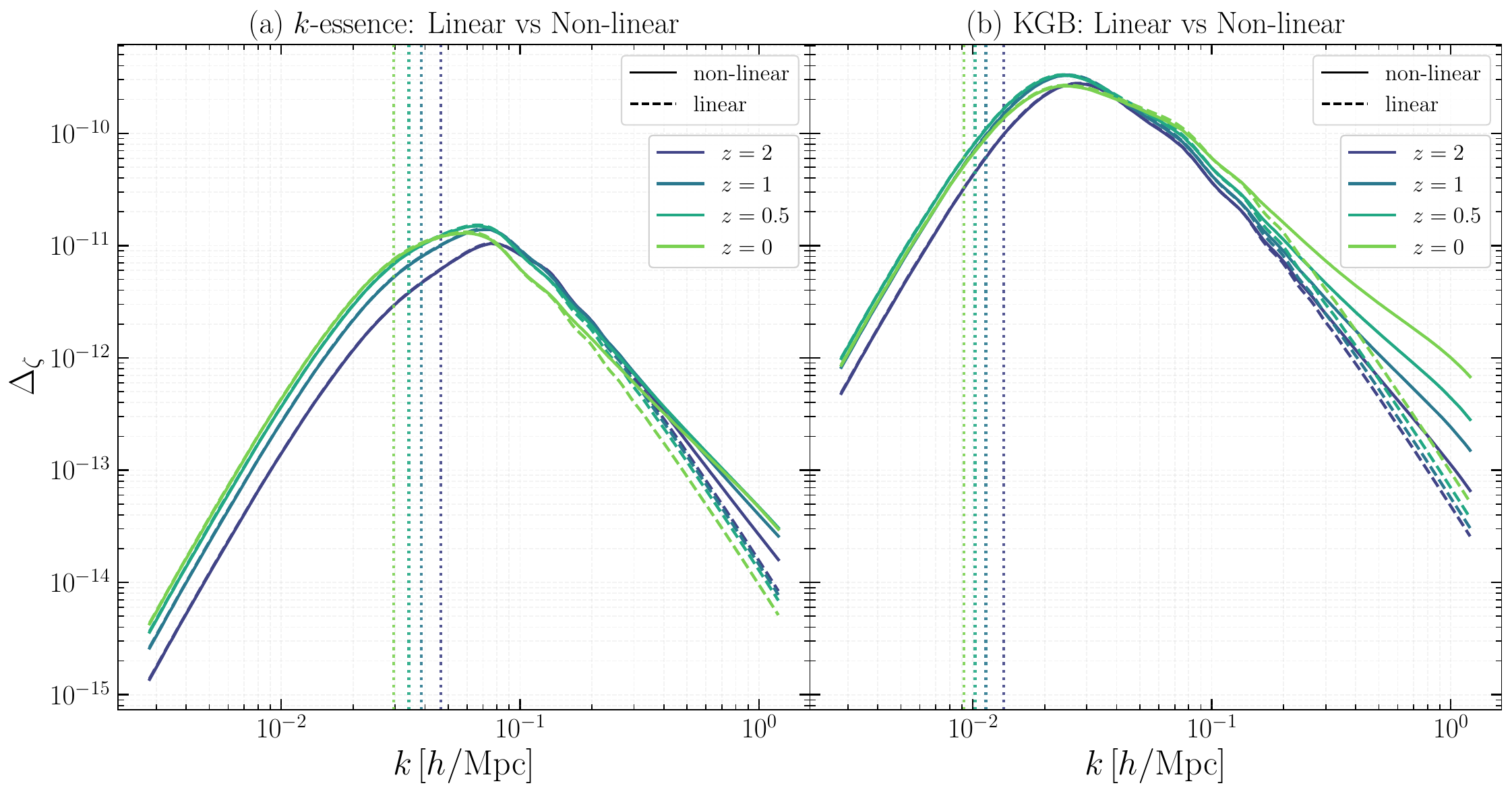}
\caption{Linear (dashed) and nonlinear (solid) $\zeta$ power spectra at redshifts $z=2,1,0.5, 0$ for (a) the $k$‑essence model and (b) the KGB model. The higher amplitude in the KGB case reflects its enhanced dynamical evolution of the scalar field perturbation driven by the braiding interaction.
}
\label{fig:zeta_power_spectrum}
\end{figure}

Subsequently, Fig. \ref{fig:zeta_power_spectrum} shows the power spectrum of 
$\zeta$.  Recall from Eq. \eqref{eq:zeta_pi} that $\zeta$ is proportional to the time derivative of the scalar field perturbation $\pi'$, thus any amplification or rapid variation in $\pi$ directly boosts $\Delta_\zeta(k)$, and indeed the KGB model produces a consistently higher $\zeta$ amplitude compared to the $k$-essence case, indicating an enhancement of the time evolution of $\pi$ when the braiding interaction is present. 
Numerically, we find that in the KGB case the $\nabla^2\pi$ dominates the dark energy contribution to $T^{0}_{0}$ in Eq. \eqref{eq:DE-Tmunu}, whereas in the $k$-essence limit the dominant term is $\zeta$ in Eq. \eqref{eq:DE-Tmunu-kess}.

\begin{figure}[t]
\centering
\includegraphics[width=1\linewidth]{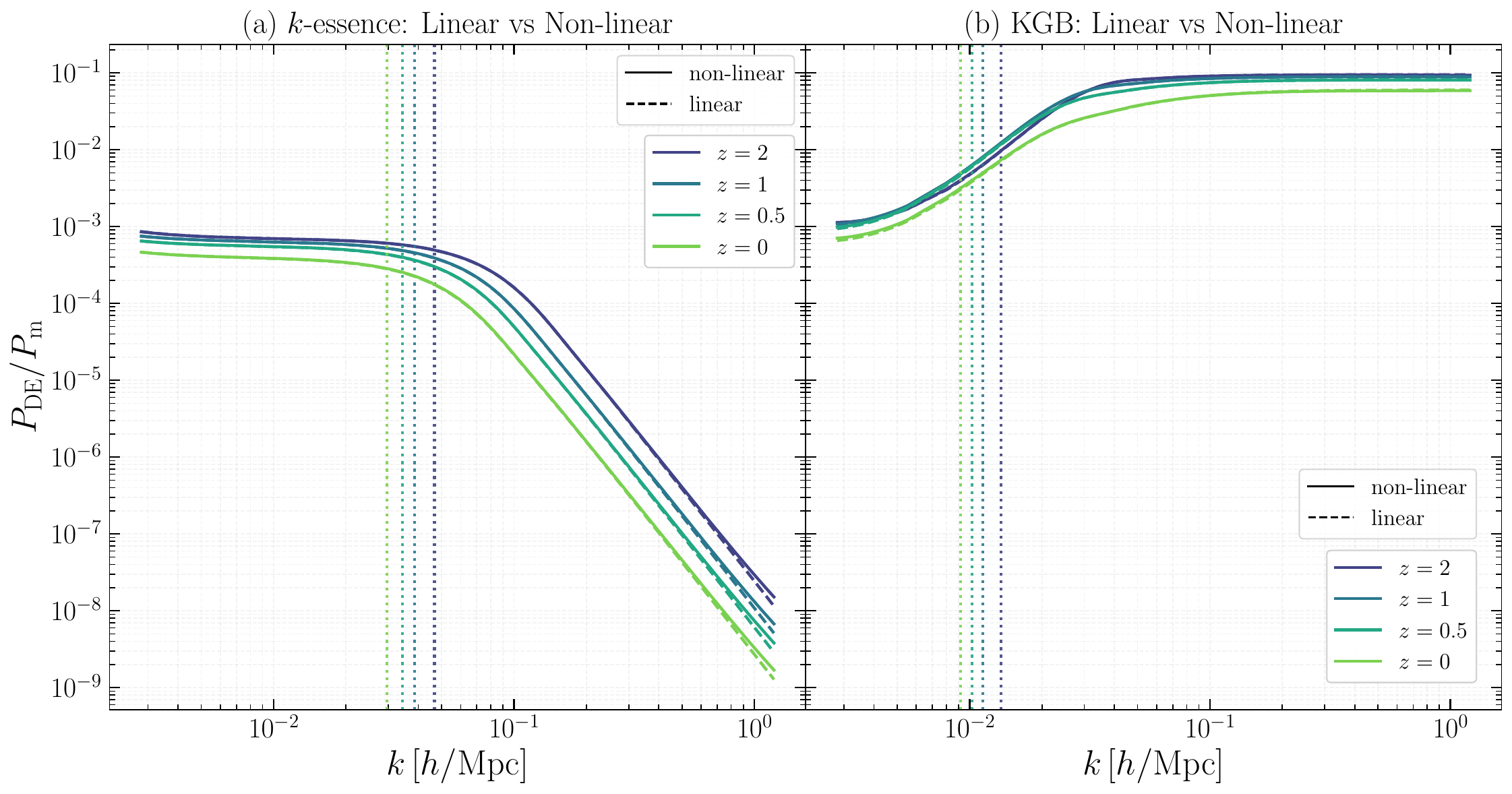}
\caption{Ratio of dark energy to dark matter power spectra at $z = 2,1,0.5, 0$. In the $k$-essence case (a) the ratio drops sharply at the sound horizon as nonlinear matter growth dominates; in the KGB model (b) braiding amplifies dark energy clustering beyond the horizon before the ratio plateaus at high $k$ where matter nonlinearity sets in.}
\label{fig:ratio_DE_m}
\end{figure}

In Fig. \ref{fig:ratio_DE_m} we display the power spectra ratio $P_{\rm DE}/P_{\rm m}$ at different redshifts. As discussed previously, in the $k$-essence case, the sound horizon is located at a scale smaller than the scale at which nonlinear matter growth begins. Consequently, while dark energy and dark matter spectra
grow almost similarly on large scales, the ratio steeply falls once modes cross the horizon and nonlinear matter growth takes over. Conversely, in the KGB model, although the time varying speed of sound pushes the horizon even farther out, the braiding term $\alpha_{\rm B}$ continually drives up dark energy fluctuations, causing $P_{\rm DE}/P_{\rm m}$ to increase across and beyond the horizon scale, and only flatten at high $k$ when nonlinear clustering of matter levels off.

\begin{figure}[t]
\centering
\includegraphics[width=1\linewidth]{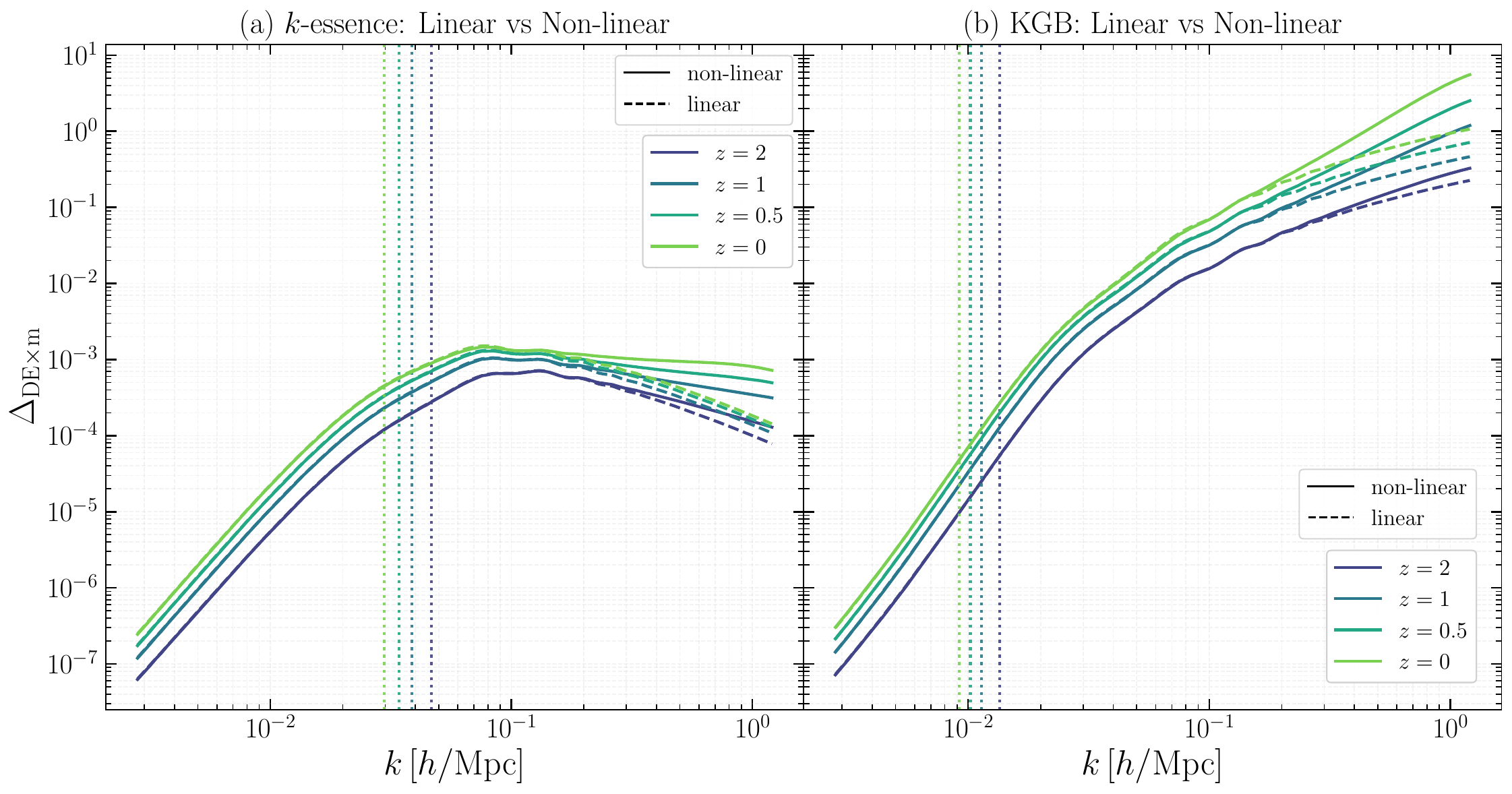}
\caption{Cross-correlation power spectra between dark energy and dark matter densities at redshifts $z = 2,1,0.5$ for (a) the $k$-essence model and (b) the KGB model. In the $k$-essence case, the spectra exhibit a turnover after crossing the sound horizon, while in the KGB model the braiding term drives a continued growth in both the linear (dashed) and nonlinear (solid) treatments, with additional amplification at small scales due to matter nonlinearities.}
\label{fig:crosscorr}
\end{figure}

Fig. \ref{fig:crosscorr} shows the cross-correlation power spectra between dark energy and dark matter density perturbations,
defined as
\begin{equation}
    \Delta_{\rm DE \times \rm m} = \frac{k^3}{2\pi^2}\,\;\big\langle \delta_{\rm DE}(\mathbf k,z)\,\delta_{\rm m}^*(\mathbf k,z)\big\rangle \, .
\end{equation}
In both models, the spectra are amplified at small scales as a result of matter nonlinearities. As expected, in the KGB model the cross-correlation power spectra continue to grow in both linear and nonlinear regimes due to the braiding term, whereas in the 
$k$-essence model the spectra exhibit a turnover after crossing the sound horizon.

We also examine the normalized cross spectrum between matter and dark energy, $f_{\rm X}(k,z)$. 
This quantity is defined as
\begin{equation}
f_{\times}(k,z)\equiv
\frac{\Delta_{\rm DE\times m}(k,z)}{\sqrt{\Delta_{\rm DE}(k,z)\Delta_{\rm m}(k,z)}} \, .
\end{equation}
By construction, $f_{\times}(k,z)$ is bounded between $-1$ and $1$. A value of $f_{\times}=1$ corresponds to perfect correlation between matter and dark energy perturbations, $f_{\times}=-1$ indicates perfect anti–correlation, and $f_{\times}=0$ shows the absence of correlation. In the linear regime and assuming adiabatic initial conditions, all perturbation fields originate from the same primordial curvature perturbation. Their subsequent evolution is governed by deterministic transfer functions, which preserve this common origin. As a result, the matter and dark energy fluctuations remain perfectly correlated and the normalized cross spectrum takes the value $f_{\times}(k,z)=1$ at all scales.

This argument no longer holds once nonlinear evolution becomes important. Mode coupling generated by nonlinear gravitational clustering induces additional correlations between different Fourier modes, and dark energy perturbations can then respond differently to the evolving matter distribution with a behavior depending in particular on the dark energy sound speed and other model parameters. As a consequence, the relation between the two fields is no longer described by a single transfer function tied to the primordial curvature perturbation, and $f_{\times}(k,z)$ can deviate from unity.

For the $k$-essence case $(\hat\alpha_{\rm K}=3000,\ \hat\alpha_{\rm B}=0)$ and the KGB benchmark $(\hat\alpha_{\rm K}=3000,\ \hat\alpha_{\rm B}=1.5)$, we find $f_{\rm X}\approx 1$ across all wavenumbers and redshifts, indicating perfect correlation between dark energy and matter clustering in both the linear and nonlinear regimes, which is a result of the large speed of sound in these cases.

\begin{figure}[t]
\centering
\includegraphics[width=1\linewidth]{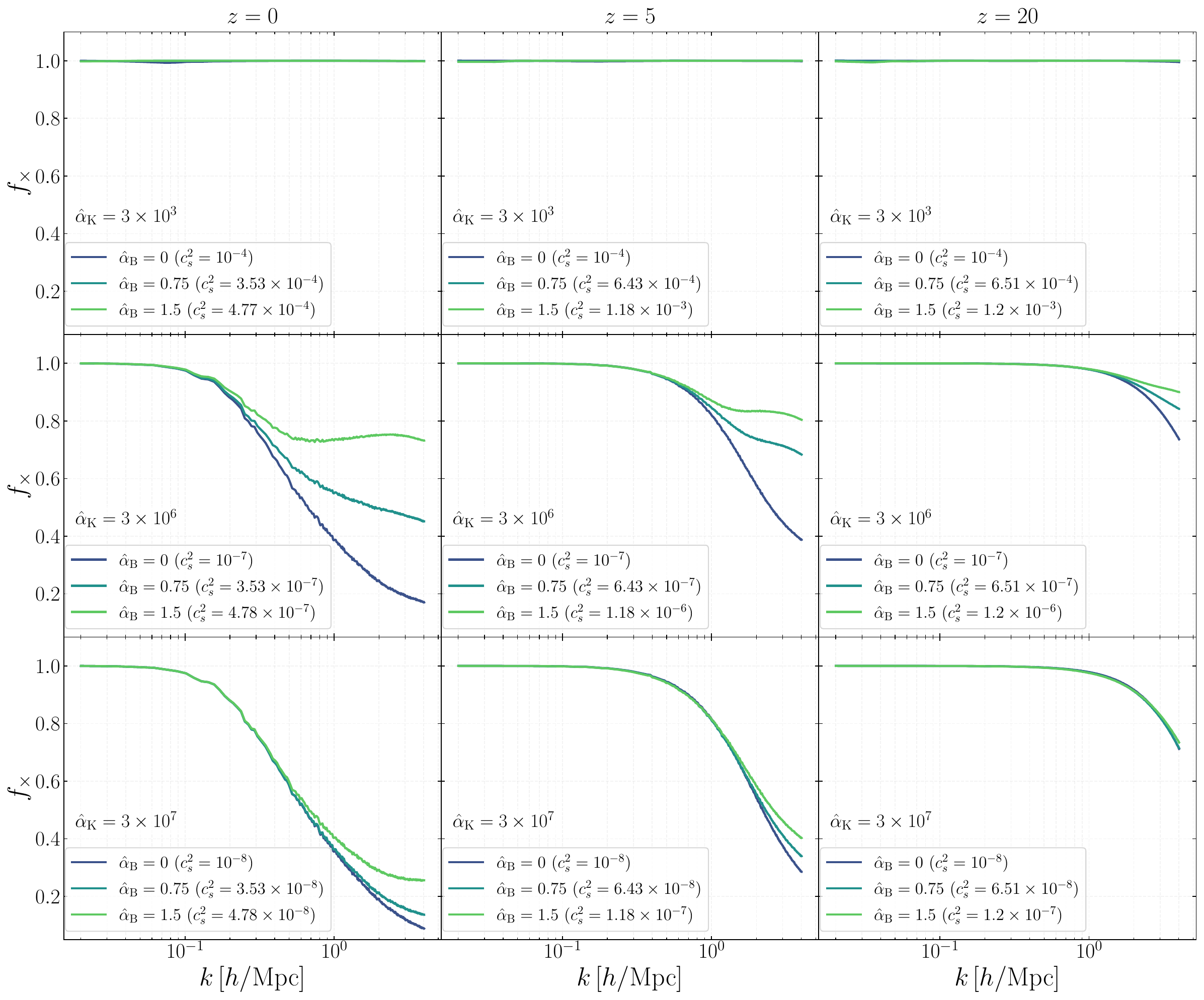}
\caption{Normalized cross-correlation power spectra between dark energy and dark matter densities at $z = 0, \, 5,\, 20$. }
\label{fig:norm-cross}
\end{figure}

To investigate this matter a bit more, in Fig. \ref{fig:norm-cross}, we display the normalized cross power spectrum at redshifts $z=0,\,5,$ and $20$, obtained from nine simulations with different choices of $\hat{\alpha}_{\rm K}$ and $\hat{\alpha}_{\rm B}$. The parameters are varied over the grid $\hat{\alpha}_{\rm K}=\{3\times10^{3},\,3\times10^{6},\,3\times10^{7}\}$ and $\hat{\alpha}_{\rm B}=\{0,\,0.75,\,1.5\}$. Each run evolves $1024^3$ particles in a simulation box of comoving size $400\,{\rm Mpc}/h$.
 The behavior of $f_{\times}$ is strongly controlled by the sound speed $c_s$, which depends primarily on $\hat{\alpha}_{\rm K}$ but is also modified by $\hat{\alpha}_{\rm B}$ through the Eq. \eqref{eq:soundspeed}.

In the top row ($\hat{\alpha}_{\rm K}=3\times10^{3}$; comparatively large $c_s^2$), $f_{\times}$ remains essentially unity at all redshifts and scales. Changing $\hat{\alpha}_{\rm B}$ in this regime slightly alters the sound speed but has little effect on $f_\times$, since $c_s^2$ is still relatively high and dark energy perturbations closely follow matter.
In the middle row ($\hat{\alpha}_{\rm K}=3\times10^{6}$; smaller $c_s^2$), clear deviations of $f_{\times}$ from unity appear at small scales and late times, both for the $k$-essence case ($\hat{\alpha}_{\rm B}=0$) and for the KGB runs ($\hat{\alpha}_{\rm B}>0$). Here, turning on braiding increases $c_s^2$ and partially restores the correlation, delaying or suppressing the drop in $f_{\times}$.
In the bottom row ($\hat{\alpha}_{\rm K}=3\times10^{7}$; lowest $c_s^2$), deviations of $f_{\times}$ from one are even stronger and extend across a broader range of scales and redshifts. In this regime, the sound speed is so small that even with nonzero $\hat{\alpha}_{\rm B}$, cannot prevent significant differences in the evolution of matter and dark energy fluctuations, and the results for 
$k$-essence and KGB appear very similar.

\subsection{Snapshots analysis}
\label{sect:snapshots}
In this section, we analyze two- and three-dimensional snapshots of dark matter and dark energy density across various redshifts in both $k$‑essence and KGB models. We specifically focus on the dark energy density contrast, $\delta = {\delta\rho/\bar{\rho}}$, varying $\alpha_{\rm B}$ parameter to investigate its influence on dark energy clustering. We perform three simulations, each using a grid of $N_{\rm grid} = 1200^3$ and a box size of 800 Mpc/$h$, yielding a spatial resolution of 0.66 Mpc/$h$. In every run, we fix $\hat\alpha_{\rm K} = 3000$ while varying $\hat\alpha_{\rm B}$ over the values $\{0,0.04,0.4\}$.
These values of $\hat\alpha_{\rm B}$ are chosen to align with the CMB+DESI+ISWL+PantheonPlus constraints of \cite{Chudaykin:2025gdn}. For this dataset, $\hat\alpha_{\rm B}=0.4$ lies within 1-$\sigma$ of the best fit, whereas $\hat\alpha_{\rm B}=0.04$ and $0$ remain allowed at the 2-$\sigma$ level.

\begin{figure}[htbp]
  \centering
  \includegraphics[width=\linewidth]{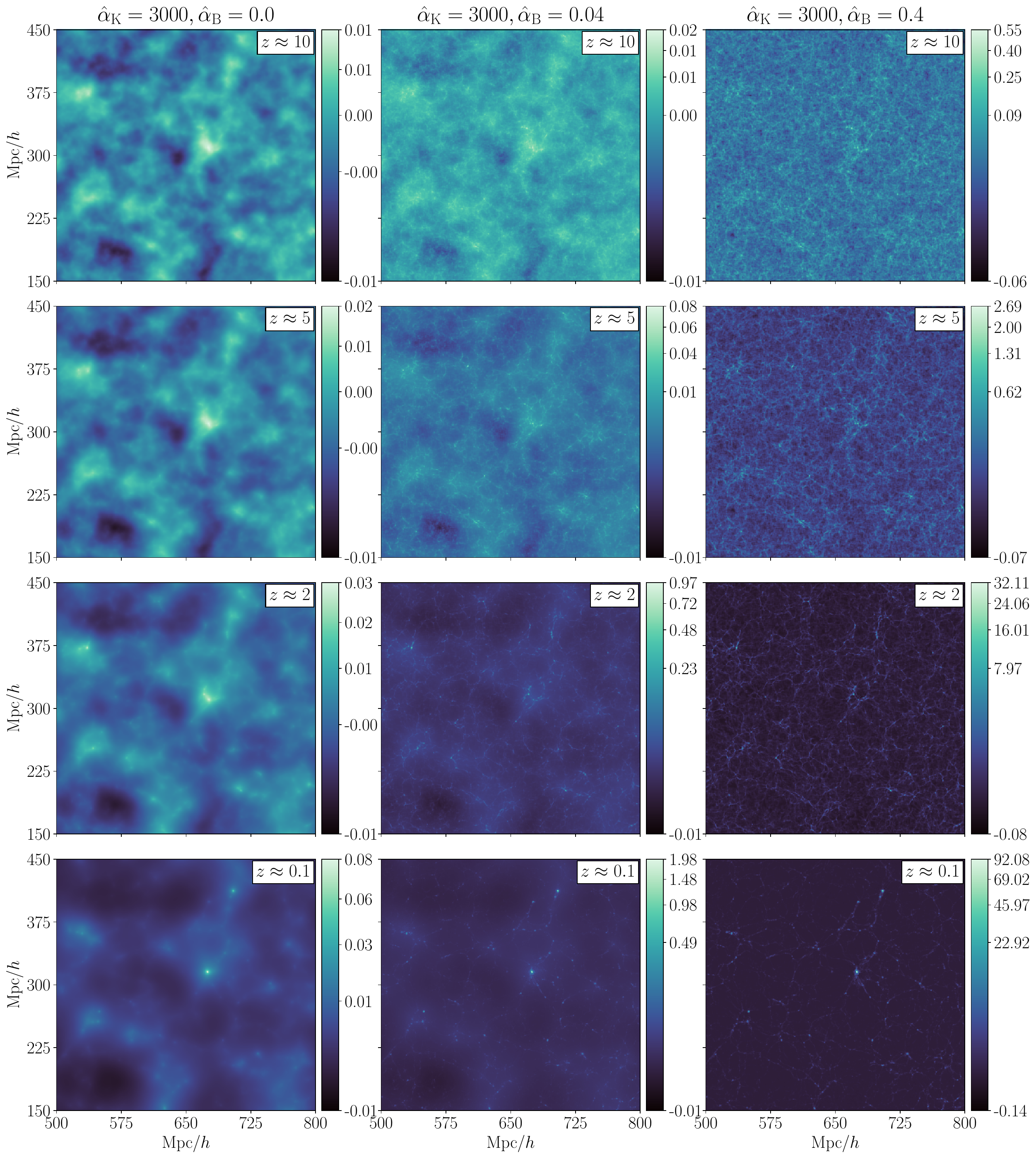}\\
  \caption{2D slices of the dark energy density contrast $\delta$, at $z=10$, 5, 2 and 0 in a subregion of the simulation box, illustrating how turning on and increasing $\hat\alpha_{\rm B}$ progressively reveals and enhances small scale structure. Colors are scaled using a power-law normalization to enhance contrast in the density field. 
For each pixel value $\delta$ within the range $[\delta_{\min}, \delta_{\max}]$, the displayed color corresponds to the normalized value
$n(\delta) = \big[(\delta - \delta_{\min})/(\delta_{\max}-\delta_{\min})\big]^{\gamma}$
which highlights structures. 
The exponent $\gamma$ controls the level of contrast enhancement.
 We use $\gamma=0.6$ for the $k$-essence case and $\gamma=0.3$ for the KGB cases to emphasize density variations.
}
  \label{fig:vertical_snapshots}
\end{figure}

Fig.~\ref{fig:vertical_snapshots} shows two‑dimensional slices of the dark energy density contrast $\delta$ 
at redshifts $z=10, 5, 2$ and 0. Each panel zooms into the subregion of the simulation box (spanning from 500 to 800 Mpc$/h$ in the $x$‑direction and from 150 to 450 Mpc$/h$ in the $y$‑direction), where dark energy overdensities are more prominent.

In the left column of Fig.~\ref{fig:vertical_snapshots}, the pure $k$‑essence case ($\hat\alpha_{\rm K}=3000$, $\hat\alpha_{\rm B}=0$) exhibits a faint, large scale clustering of the dark energy density contrast across all redshifts (top to bottom: $z=10$, 5, 2, 0). This baseline clustering remains relatively smooth, with only subtle variations indicating the onset of structure formation at low redshifts. Therefore, for this case, the dark energy density contrast remains always in a quasi-linear regime.
When $\hat\alpha_{\rm B}$ is raised to 0.04 (middle column), a distinct substructure emerges within the broader $k$‑essence pattern. At $z=10$, localized overdense regions appear as slight intensity enhancements in the density field, but by $z=5$ and $z=2$ they evolve into compact nonlinear structures embedded within the larger contrast areas. By $z=0$, these secondary enhancements have sharpened significantly, indicating that even a modest braiding term can boost small scale dark energy clustering.

In the right column ($\hat\alpha_{\rm B}=0.4$), the braiding induced overdensities become even more pronounced. Already at $z=10$, we can see a huge enhancement of dark energy clustering compared to the two previous cases, and at $z=0$ they appear as the brightest, most concentrated and highly nonlinear features. This indicates that increasing $\hat\alpha_{\rm B}$ and progressing to lower redshifts both amplify and sharpen the inner clustering driven by the braiding parameter, on top of the underlying $k$‑essence clustering.

\begin{figure}[t]
\centering
\includegraphics[width=1\linewidth]{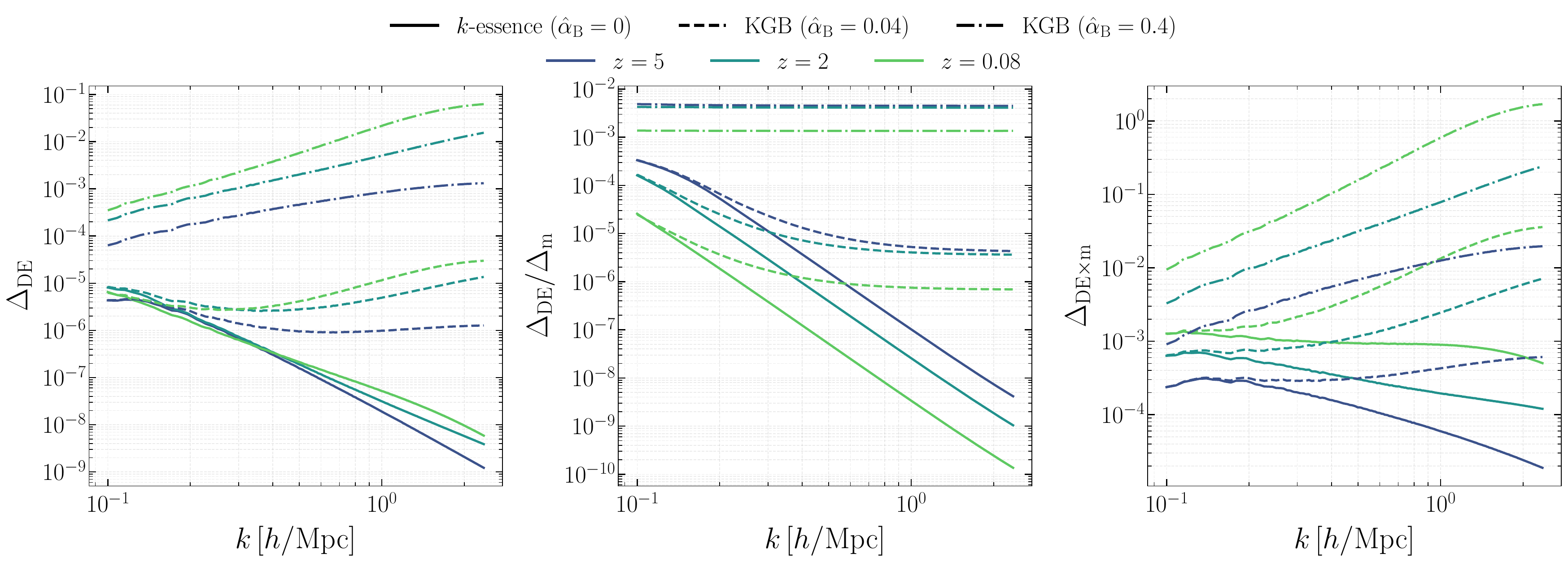}
\caption{Power spectra of dark energy $\Delta_{\rm DE}$, dark energy-matter ratio $\Delta_{\rm DE}/ \Delta_{\rm m}$ and  dark energy-matter cross correlation $\Delta_{\rm DE \times m}$ at  $z = 5, \, 2, \, 0.08$. }
\label{fig:snap_power}
\end{figure}

For completeness, the power spectra corresponding to these snapshots (at $z=5, 2, 0.08$) are shown in Fig. \ref{fig:snap_power}, which displays $\Delta_{\rm DE}$ (left), the ratio $\Delta_{\rm DE}/\Delta_{\rm m}$ (middle), and the dark energy–matter cross power $\Delta_{\rm DE\times m}$ (right).  Across all three panels, turning on braiding parameter raises the dark energy signal relative to the $k$-essence case and tightens its connection to the matter field: at fixed redshift the separation between the curves grows toward smaller scales, and at fixed scale it increases toward lower redshift. The $\hat\alpha_{\rm B}=0.04$ model already shows a clear enhancement over $k$-essence, while $\hat\alpha_{\rm B}=0.4$ noticeably boosts $\Delta_{\rm DE}$ and $\Delta_{\rm DE}/\Delta_{\rm m}$, and strengthens $\Delta_{\rm DE\times m}$, indicating more pronounced and more correlated dark energy clustering that becomes most evident at late times and on mildly nonlinear scales.

\begin{figure}[ht!]
  \centering
  \begin{overpic}[width=\linewidth]
  {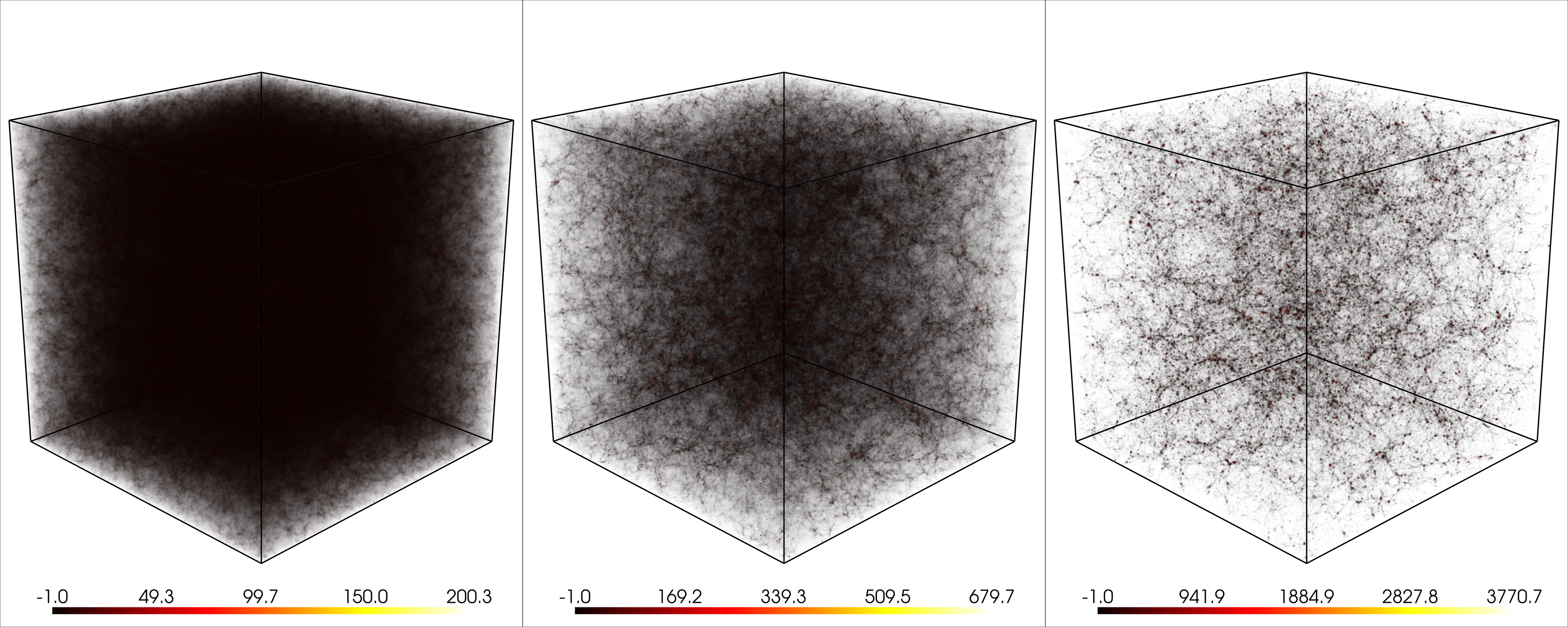}
  \put(0,41.5){\small\textbf{Matter:}}
    \put(1,37.5){\small\textbf{$z \approx 5$}}
    \put(34.2,37.5){\small\textbf{$z \approx 2$}}
    \put(67.4,37.5){\small\textbf{$z \approx 0.1$}}
  \end{overpic}\\[5.5ex]
  \begin{overpic}[width=\linewidth]
  {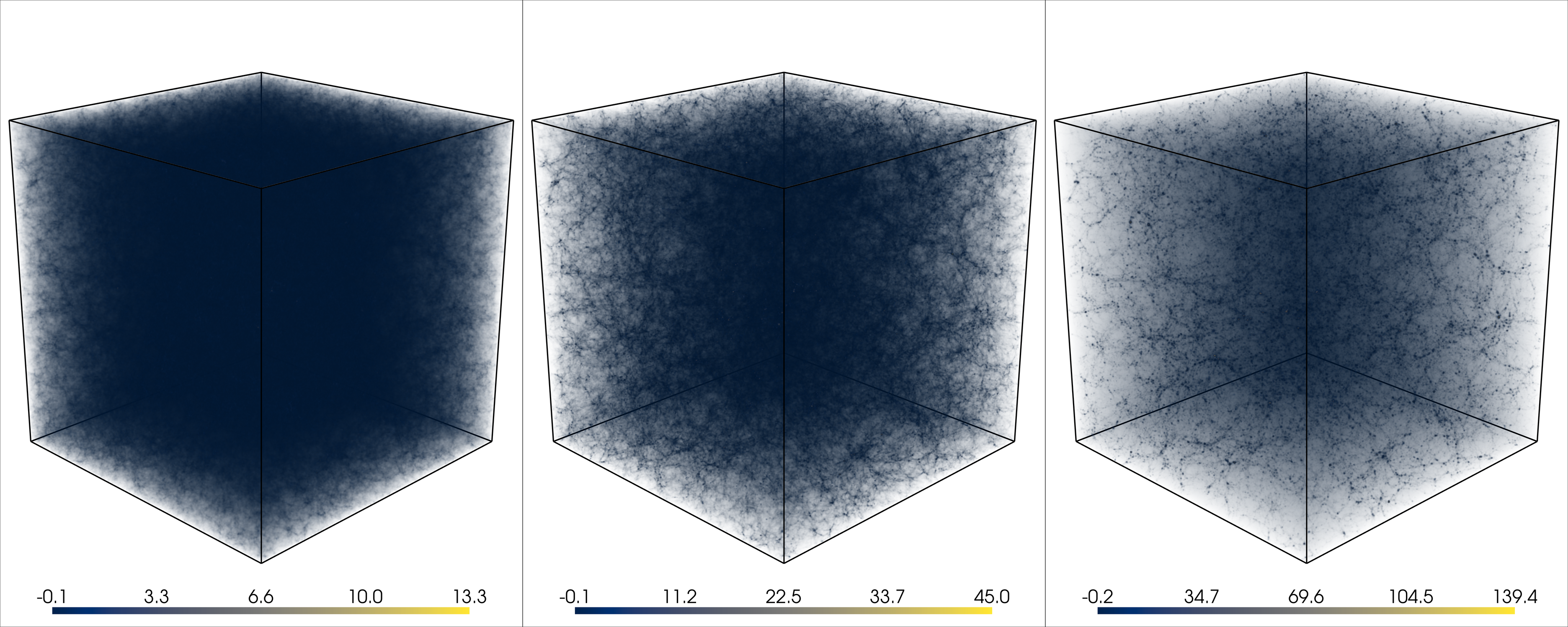}
  \put(0,41.5){\small\textbf{KGB: ($\hat{\alpha}_{\rm K} = 3000$, $\hat{\alpha}_{\rm B} = 0.4$ )}}
    \put(1,37.5){\small\textbf{$z \approx 5$}}
    \put(34.2,37.5){\small\textbf{$z \approx 2$}}
    \put(67.4,37.5){\small\textbf{$z \approx 0.1$}}
  \end{overpic}\\[5.5ex]
    \begin{overpic}[width=\linewidth]{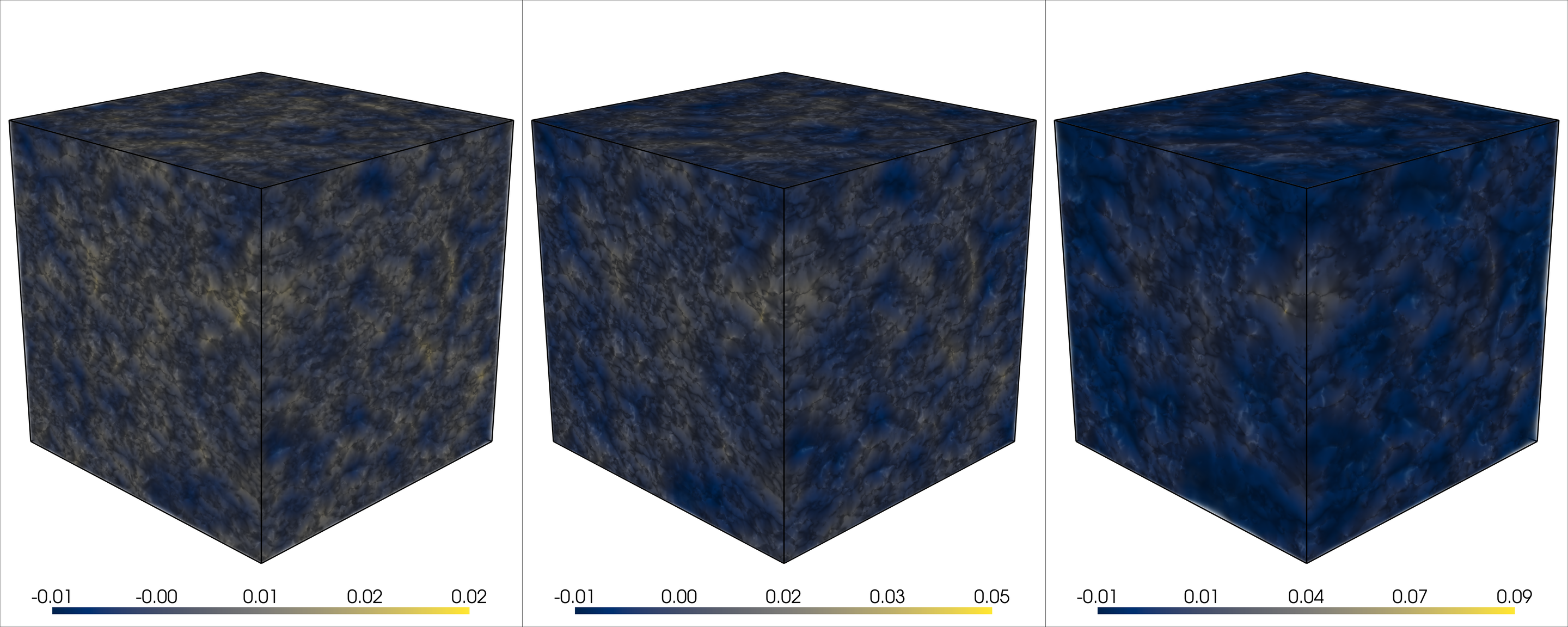}
      \put(0,41.5){\small\textbf{$k$-essence: ($\hat{\alpha}_{\rm K} = 3000$, $\hat{\alpha}_{\rm B} = 0$ )}}
    \put(1,37.5){\small\textbf{$z \approx 5$}}
    \put(34.2,37.5){\small\textbf{$z \approx 2$}}
    \put(67.4,37.5){\small\textbf{$z \approx 0.1$}}
  \end{overpic}
  \caption{3D snapshots of matter and dark energy density contrast in both KGB and $k$-essence model. The color bars show the density contrast, $\delta$.}
  \label{fig:darkenergy_box}
\end{figure}
For visualization purposes, Fig. \ref{fig:darkenergy_box} shows 3D rendering of the density contrast in matter and dark energy for the KGB model ($\hat\alpha_{\rm K}=3000$, $\hat\alpha_{\rm B}=0.4$) and the $k$-essence model ($\hat\alpha_{\rm K}=3000$, $\hat\alpha_{\rm B}=0$) at redshifts $z=5$, $2$, and $0$, produced with PyVista \cite{sullivan2019pyvista}. These snapshots are generated from the same simulations used in Fig. \ref{fig:vertical_snapshots}, but restricted to a box size ranging from 350 to 750 Mpc/$h$ along each dimension.
These 3D renderings make it clear that with $\hat\alpha_{\rm B}=0.4$ dark energy in the KGB model closely follows the filamentary matter distribution, whereas in the $k$‑essence case ($\hat\alpha_{\rm B}=0$) it remains a single, diffuse overdensity with only weak, large scale clustering at low redshifts.

\newpage
\section{Conclusion}
\label{sec:conclusions}
In this work we introduced \kgb, an extension of the fully relativistic $N$-body code of \kev that implements kinetic gravity braiding dark energy through the effective field theory $\alpha$-parametrization. Starting from the covariant Horndeski action and reformulating it in terms of ${\alpha_{\rm K},\alpha_{\rm B}}$, we derived and implemented the linearized dark energy stress–energy tensor and scalar field equation in the \kev/\gev framework, leaving the nonlinear dark energy self-interactions to future work. Our implementation reproduces \kev outputs in the $k$-essence limit ($\hat\alpha_{\rm B}=0$), and agrees at the sub-percent level against linear \hiclass predictions for matter and power spectrum 
on large scales at low redshift and on all scales at high redshift. 

Turning on braiding ($\hat\alpha_{\rm B}\neq0$) produces scale- and redshift-dependent departures from the $k$-essence limit. The matter power spectrum shows a clear departure from $k$-essence in linear theory, and nonlinear evolution further amplifies this difference, showing that KGB effects are amplified on quasi-linear and nonlinear scales. The gravitational potential spectrum is even more sensitive: braiding enhances $\Delta_\Phi$ across a broad range of scales, implying that lensing/ISW-type observables should constrain KGB more powerfully than matter sector probes. 

From the dark energy side, both spectra and snapshots show that KGB clusters significantly more than $k$-essence. Inside and outside the sound horizon, $\alpha_{\rm B}$ sustains growth of $\delta_{\rm DE}$ -- in contrast to the decay seen for $k$-essence -- and in our simulations this drives a visibly more inhomogeneous dark energy field. We further show, both at linear order and in our relativistic 
$N$-body simulations, a characteristic feature of KGB dynamics: the scalar field perturbation $\pi$ undergoes a zero-crossing (sign flip) inside the horizon, which manifests as a turnover in the corresponding power spectra; a behavior which is absent in the $k$-essence limit.  In real space, 2D/3D snapshots confirm progressively stronger small scale dark energy clustering as $\hat\alpha_{\rm B}$ increases, with the KGB field tracing the filamentary matter web far more closely than in $k$-essence. 

Practically, \kgb is ready for forward-modelling pipelines: combined with \gev{’s} light-cone capability, one can build observables 
directly from simulations, enabling survey-level comparisons beyond linear theory, which would be crucial in the next generation cosmological surveys such as  \textit{Euclid} and DESI, 
where precision modelling of late-time clustering is essential. Building on this, an emulator trained on \kgb outputs, similar to \cite{Nouri-Zonoz:2024dph}, can be embedded in MCMC analyses to tighten constraints on $(\alpha_{\rm K},\alpha_{\rm B})$ over relevant prior volumes.

Going forward, extending \kgb to evolve the full nonlinear dark energy sector will allow us to capture higher-order effects and study the stability domain of KGB models in a relativistic $N$-body setting.

\acknowledgments
We are grateful to Julian Adamek, Alexander Vikman, and Abdolali Banihashemi for valuable discussions and insightful comments.
Ahmad Nouri-Zonoz and Martin Kunz acknowledge funding from the Swiss National Science Foundation.
Farbod Hassani acknowledges the Research Council of Norway and the resources provided by 
UNINETT Sigma2 -- the National Infrastructure for High Performance Computing and 
Data Storage in Norway. 
Ahmad Nouri-Zonoz would like to thank the Vahabzadeh Foundation for their continuous financial support. Farbod Hassani acknowledges funding from the Research Council of Norway (Young Talent Grant, Project No. 345334).
Emilio Bellini is supported by the European Union’s Horizon Europe research and innovation programme under the Marie Sklodowska-Curie Postdoctoral Fellowship Programme, SMASH co-funded under the grant agreement No.~101081355.
The simulations in this work were performed on resources provided by
Sigma2 -- the National Infrastructure for High-Performance Computing and
Data Storage in Norway.

\newpage
\appendix
\section{$k$-essence limit}
\label{app:k-esselim} 

In $k$-essence where $\alpha_\textrm{B} = 0$, the stress-energy  tensor and the equation of motion take the form as Eqs. \eqref{eq:DE-Tmunu-kess} and \eqref{eq:linearEOM-kess}, respectively as shown in Table \ref{tab:kess-equations}.
\begin{table}[htbp]
\centering
\hspace{-0.5cm}
\caption{Linear equations for stress-energy tensor and scalar field equation in $k$-essence model. 
}
\label{tab:kess-equations}
\begin{tcolorbox}[colframe=black, colback=white!10, boxrule=0.5mm, sharp corners=south, 
  title=\textbf{$k$-essence equations}, halign title=center]
\textbf{Stress-energy tensor:}
\begin{equation}
\begin{aligned}
T^0_0 &= - \bar{\rho}_\phi  + \frac{M_\textrm{Pl}{}^2}{a^2} \bigg\{  -\alpha_\textrm{K} \mathcal{H}^2 \zeta  +\Big[ \frac{a^2}{M_\textrm{Pl}{}^2} (\bar{\rho}_\phi  + \bar{P}_\phi )\Big]3 \mathcal{H} \pi \bigg\} \, ,\\[8pt] 
T^0_i &=-(\bar{\rho}_\phi+\bar{P}_\phi)\partial_{i}\pi \, ,\\[8pt]
T^i_j &= \big(\bar{P}_\phi + \pi  \bar{P}_{\phi}^{\prime}\big) \delta^i_j  +\zeta (\bar{\rho}_\phi  + \bar{P}_\phi)\delta^i_j\, .
\end{aligned}\label{eq:DE-Tmunu-kess}
\end{equation}

\textbf{Equation of motion:}
\begin{equation}
 \textcolor{BrickRed}{A_{\zeta'}} \zeta' + \textcolor{BrickRed}{A_{\Phi'}} \Phi' + \textcolor{BrickRed}{A_{\nabla^2 \pi}} \nabla^2 \pi + \textcolor{BrickRed}{A_{\Psi}} \Psi + \textcolor{BrickRed}{A_{\pi}} \pi + \textcolor{BrickRed}{A_{\zeta}} \zeta = 0 \, ,
\label{eq:linearEOM-kess}
\end{equation}
where
\begin{align*}
\textcolor{BrickRed}{A_{\zeta'}} &= \alpha_\textrm{K}~,~
   \textcolor{BrickRed}{A_{\Phi'}} =  \frac{a^2}{M_\text{Pl}^2} \left(-\frac{3 \bar{\rho}_\phi}{\mathcal{H}^2} - \frac{3 \bar{P}_\phi}{\mathcal{H}^2}\right)~,~
   \textcolor{BrickRed}{A_{\nabla^2 \pi}} =  \frac{a^2}{M_\text{Pl}^2} \left(-\frac{\bar{\rho}_\phi}{\mathcal{H}^2} - \frac{\bar{P}_\phi}{\mathcal{H}^2}\right) \, , \\
   \textcolor{BrickRed}{A_{\Psi}} &= \frac{a^2}{M_\text{Pl}^2} \left(-\frac{3 \bar{\rho}_\phi}{\mathcal{H}} - \frac{3 \bar{P}_\phi}{\mathcal{H}}\right)~,~
   \textcolor{BrickRed}{A_{\pi}}= 
    \frac{a^2}{M_\text{Pl}^2} \left(3 \bar{\rho}_\phi - \frac{3 \mathcal{H}' \bar{\rho}_\phi}{\mathcal{H}^2} + 3 \bar{P}_\phi - \frac{3 \mathcal{H}' \bar{P}_\phi}{\mathcal{H}^2}\right)\, , \\
   \textcolor{BrickRed}{A_{\zeta}} &= \alpha_\textrm{K}' 
   + \alpha_\textrm{K} \left(\mathcal{H} + \frac{2 \mathcal{H}'}{\mathcal{H}}\right) \, .
\end{align*}
\end{tcolorbox}
\end{table}

To check the consistency of these equation with \cite{Hassani:2019lmy}, we can map them to fluid language using the following relations
\begin{equation}
    w = \frac{\bar{P}_\phi}{\bar{\rho}_\phi}, \quad c_s^2 =  \frac{\left(\bar{\rho}_\phi+\bar{P}_\phi\right) a^2}{M_\text{Pl}^2\alpha_k \mathcal{H}^2}, \quad c_a^2 = \frac{\bar{P}_\phi^{\prime}}{\bar{\rho}_\phi^{\prime}} \, ,
    \label{eq:fluidlang}
\end{equation}
where $w$, $ c_s^2$, and $c_a^2$ represent the equation of state parameter, the sound speed squared, and the adiabatic sound speed squared, respectively. Using these we can rewrite the $k$-essence stress-energy tensor and equation of motion as
\begin{align}
T^0_0  &= - \bar{\rho}_\phi + \frac{\bar{P}_\phi+\bar{\rho}_\phi}{c_s^2} \big(3 c_s^2 \mathcal{H}\pi - \zeta \big) \, ,\\
T^0_i &= -(\bar{\rho}_\phi+\bar{P}_\phi)\partial _{i}\pi \, ,\\
T^i_j &= \bar{P}_\phi \delta^i_j-(\bar{\rho}_\phi+\bar{P}_\phi)\left[3 c_{a}^{2}\mathcal{H}  \pi  - \zeta  \right]\delta^i_j \, ,
\end{align}
and 
\begin{align}
& \zeta ^{\prime}  =3 w \mathcal{H}  \zeta -3 c_s^2( \mathcal{H}^2\pi- \mathcal{H}\Psi  - \mathcal{H}^{\prime}\pi -\Phi^{\prime})  + c_s^2 \nabla ^{2}\pi, \nonumber \\ & \zeta= \pi' + \mathcal H \pi -\Psi,
\end{align}
which matches the equations in \cite{Hassani:2019lmy}.

\begin{figure}[t]
\centering
\includegraphics[width=1\linewidth]{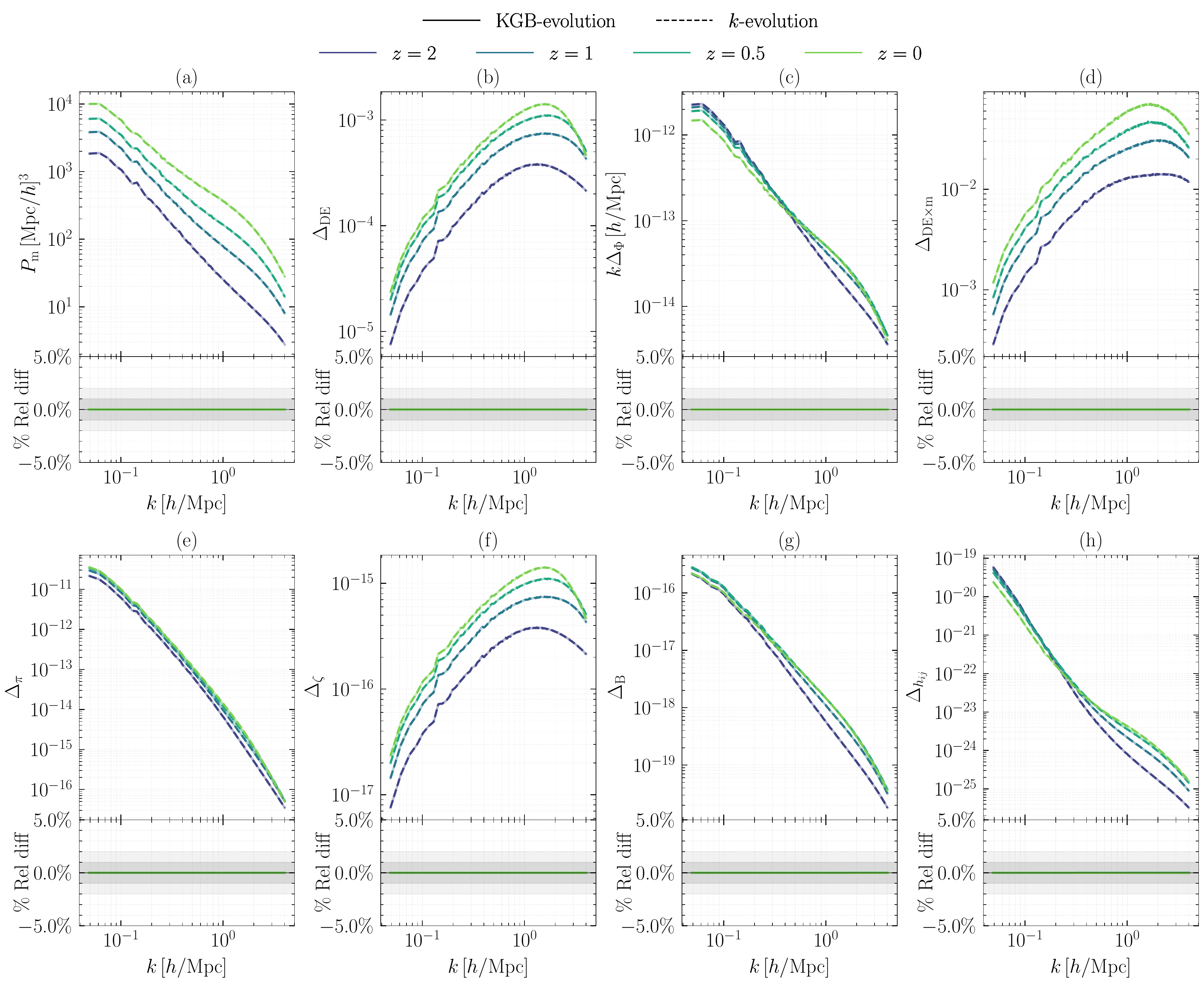}
\caption{ Comparison of \kgb (solid lines) in the $k$-essence limit ($\hat{\alpha}_{\rm B}=0$)  with the $k$-evolution code (dashed lines). The panels show power spectra of: (a) matter density, (b) dark energy density, (c) gravitational potential, (d) matter- dark energy cross correlation, (e) scalar field dark energy perturbation, (f) $\zeta$, (g) vector perturbations, and (h) tensor perturbations. Colors indicate redshifts $z=\{0, \, 0.5, \, 1, \, 2\}$ The subplots show the relative difference between the two codes with grey bands marking $\pm1\%$ and $\pm2\%$.}
\label{fig:KGBvskev}
\end{figure}

In Fig. \ref{fig:KGBvskev} we show that, in the $k$-essence limit ($\hat{\alpha}_{\rm B}=0$), \kgb code, implemented in $\alpha$-parameterization is fully compatible with the $k$-evolution code,  which is formulated in terms of the fluid variables defined in Eq \eqref{eq:fluidlang} for the stress–energy tensor and the dark energy scalar field equation. Across all redshifts and the full $k$-range shown, the spectra agree very well with relative differences curves remaining at the sub-percent level, confirming the consistency of \kgb code at its $k$-essence limit with \kev.

\section{Numerical Implementation}
\label{Numerical Implementation}
\subsection{Scalar field equation}
We apply the Newton-Stormer-Verlet-leapfrog method \cite{Hairer_Lubich_Wanner_2003} to solve the partial differential equation for $\zeta$ and $\pi$ on the lattice
\begin{equation}
    \zeta_{\mathrm{i}, \mathrm{j}, \mathrm{k}}^{\mathrm{n}+\frac{1}{2}}=\zeta_{\mathrm{i}, \mathrm{j}, \mathrm{k}}^{\mathrm{n}-\frac{1}{2}}+\zeta_{\mathrm{i}, \mathrm{j}, \mathrm{k}}^{\prime \mathrm{n}} \Delta \tau \, .
    \label{zetaleap}
\end{equation}
Here, the upper index indicates the discrete time steps, the lower indices specify the lattice‐point coordinates, and $\Delta\tau$ denotes the conformal time step-size. From \eqref{eq:linearEOM}, one can compute the discrete time‐derivative of $\zeta$ at time step $n$ and lattice site $(i,j,k)$
\begin{equation}
      \zeta_{\mathrm{i}, \mathrm{j}, \mathrm{k}}^{\prime \mathrm{n}} = - \frac{1}{\textcolor{BrickRed}{A_{\zeta'}}} \bigg[  \textcolor{BrickRed}{A_{\zeta}} \zeta_{\mathrm{i}, \mathrm{j}, \mathrm{k}}^{\mathrm{n}} + \textcolor{BrickRed}{A_{\pi}} \pi_{\mathrm{i}, \mathrm{j}, \mathrm{k}}^{\mathrm{n}} + \textcolor{BrickRed}{A_{\Psi}} \Psi_{\mathrm{i}, \mathrm{j}, \mathrm{k}}^{\mathrm{n}}  + \textcolor{BrickRed}{A_{\Phi'}} \Phi_{\mathrm{i}, \mathrm{j}, \mathrm{k}}^{\prime \mathrm{n}}  + \textcolor{BrickRed}{A_{\nabla^2 \pi}} \nabla^2 \pi + \textcolor{BrickRed}{A_{\nabla^2 \Psi}} \nabla^2 \Psi   + \textcolor{BrickRed}{A_{\delta P_\textrm{m}}} \delta P_{\mathrm{i}, \mathrm{j}, \mathrm{k}}^{\mathrm{n}}  \bigg] \, ,
      \label{zetanprime}
\end{equation}
where the Laplacian $\nabla^2\pi$ and $\nabla^2\Psi$ are evaluated with the standard six-point stencil as follow
\begin{equation}
\begin{aligned}
    \nabla^2 \pi &= \frac{\pi_{\mathrm{i}-1, \mathrm{j}, \mathrm{k}}^{\mathrm{n}}+\pi_{\mathrm{i}+1,\mathrm{j}, \mathrm{k}}^{\mathrm{n}}+\pi_{\mathrm{i}, \mathrm{j}-1, \mathrm{k}}^{\mathrm{n}}+\pi_{\mathrm{i}, \mathrm{j}+1, \mathrm{k}}^{\mathrm{n}}+\pi_{\mathrm{i}, \mathrm{j}, \mathrm{k}-1}^{\mathrm{n}}+\pi_{\mathrm{i}, \mathrm{j}, \mathrm{k}+1}^{\mathrm{n}}-6 \pi_{\mathrm{i}, \mathrm{j}, \mathrm{k}}^{\mathrm{n}}}{\Delta \mathrm{x}^2} \, , \\
    ~\\
    \nabla^2 \Psi &= \frac{\Psi_{\mathrm{i}-1, \mathrm{j}, \mathrm{k}}^{\mathrm{n}}+\Psi_{\mathrm{i}+1,\mathrm{j}, \mathrm{k}}^{\mathrm{n}}+\Psi_{\mathrm{i}, \mathrm{j}-1, \mathrm{k}}^{\mathrm{n}}+\Psi_{\mathrm{i}, \mathrm{j}+1, \mathrm{k}}^{\mathrm{n}}+\Psi_{\mathrm{i}, \mathrm{j}, \mathrm{k}-1}^{\mathrm{n}}+\Psi_{\mathrm{i}, \mathrm{j}, \mathrm{k}+1}^{\mathrm{n}}-6 \Psi_{\mathrm{i}, \mathrm{j}, \mathrm{k}}^{\mathrm{n}}}{\Delta \mathrm{x}^2} \, ,
    \end{aligned}
\end{equation}
with $\Delta {\rm x}$ being the comoving spatial resolution.
The 
field $\zeta_{\mathrm{i}, \mathrm{j}, \mathrm{k}}^{ \mathrm{n}}$ at integer time step, can be obtained by taking the mean of its forward and backward half–step values
\begin{equation}
    \zeta_{\mathrm{i}, \mathrm{j}, \mathrm{k}}^{ \mathrm{n}}  = \frac{\zeta_{\mathrm{i}, \mathrm{j}, \mathrm{k}}^{ \mathrm{n}+\frac{1}{2}} + \zeta_{\mathrm{i}, \mathrm{j}, \mathrm{k}}^{ \mathrm{n}-\frac{1}{2}}}{2}.
    \label{zetan}
\end{equation}
Putting \eqref{zetan} into \eqref{zetanprime} and then \eqref{zetanprime} into \eqref{zetaleap}, we can write
\begin{align}
    \zeta_{\mathrm{i}, \mathrm{j}, \mathrm{k}}^{\mathrm{n}+\frac{1}{2}}=\zeta_{\mathrm{i}, \mathrm{j}, \mathrm{k}}^{\mathrm{n}-\frac{1}{2}} - \frac{1}{\textcolor{BrickRed}{A_{\zeta'}}} \Bigg(  \textcolor{BrickRed}{A_{\zeta}} \left[\frac{\zeta_{\mathrm{i}, \mathrm{j}, \mathrm{k}}^{ \mathrm{n}+\frac{1}{2}} + \zeta_{\mathrm{i}, \mathrm{j}, \mathrm{k}}^{ \mathrm{n}-\frac{1}{2}}}{2}\right]+ \textcolor{BrickRed}{A_{\pi}} \pi_{\mathrm{i}, \mathrm{j}, \mathrm{k}}^{\mathrm{n}} + \textcolor{BrickRed}{A_{\Psi}} \Psi_{\mathrm{i}, \mathrm{j}, \mathrm{k}}^{\mathrm{n}}  + \textcolor{BrickRed}{A_{\Phi'}} \Phi_{\mathrm{i}, \mathrm{j}, \mathrm{k}}^{\prime \mathrm{n}}\\ \nonumber  + \textcolor{BrickRed}{A_{\nabla^2 \pi}} \nabla^2 \pi + \textcolor{BrickRed}{A_{\nabla^2 \Psi}} \nabla^2 \Psi   + \textcolor{BrickRed}{A_{\delta P_\textrm{m}}} \delta P_{\mathrm{i}, \mathrm{j}, \mathrm{k}}^{\mathrm{n}}  \Bigg) \Delta \tau \, .
\end{align}
Similarly, for $\pi$ we can write
\begin{align}
    \pi^{\mathrm{n}+1}_{\mathrm{i},\mathrm{j},\mathrm{k}} &= \pi^\mathrm{n}_{\mathrm{i},\mathrm{j},\mathrm{k}} + \pi'^{ \mathrm{n}+\frac{1}{2}}_{\mathrm{i},\mathrm{j},\mathrm{k}} \Delta \tau \nonumber\\
    &= \pi^\mathrm{n}_{\mathrm{i},\mathrm{j},\mathrm{k}} +\Big[\zeta^{\mathrm{n}+\frac12}+\Psi^{\mathrm{n}+\frac12}-\mathcal H\frac{\pi^{\mathrm{n}+1}+\pi^{\mathrm{n}}}{2}\Big]\Delta \tau \,,
    \end{align}
where we replaced $\pi'^{ \mathrm{n}+\frac{1}{2}}_{\mathrm{i},\mathrm{j},\mathrm{k}}$ using \eqref{eq:zeta_pi}, and used $\pi^{\mathrm{n}+\frac{1}{2}} = \pi^{\mathrm{n}+1}+\pi^{\mathrm{n}}/2$ \,.

Isolating $\zeta_{\mathrm{i}, \mathrm{j}, \mathrm{k}}^{\mathrm{n}+\frac{1}{2}}$ and $\pi^{\mathrm{n}+1}_{\mathrm{i},\mathrm{j},\mathrm{k}}$ leads to the equations of Table \ref{tab:zeta_pi_imp}. These equations are implemented in the functions \texttt{update\_zeta} and \texttt{update\_pi} respectively inside \texttt{gevolution.hpp} file.
\begin{table}[htbp]
    \centering
    \hspace{-0.5cm}
    \caption{Evolution equations for $\zeta$ and $\pi$ implemented in the code.}
\begin{tcolorbox}[
  colframe=black,
  colback=white!10!white,
  boxrule=0.5mm,
  title=\textbf{Equation of motion},
  halign title=center,
  width=\textwidth
]
\begin{align}
\zeta_{\mathrm{i},\mathrm{j},\mathrm{k}}^{\mathrm{n}+\tfrac{1}{2}}
&= \frac{1}{1 + \bigl(\tfrac{\textcolor{BrickRed}{A_{\zeta}}}{\textcolor{BrickRed}{A_{\zeta'}}}\bigr)\,\tfrac{\Delta \tau}{2}}
\Biggl[
\zeta_{\mathrm{i},\mathrm{j},\mathrm{k}}^{\mathrm{n}-\tfrac{1}{2}}
- \frac{\Delta \tau}{\textcolor{BrickRed}{A_{\zeta'}}}\,\Bigl(
    \tfrac{\textcolor{BrickRed}{A_{\zeta}}}{2}\,\zeta_{\mathrm{i},\mathrm{j},\mathrm{k}}^{\mathrm{n}-\tfrac{1}{2}}
  + \textcolor{BrickRed}{A_{\pi}}\,\pi_{\mathrm{i},\mathrm{j},\mathrm{k}}^\mathrm{n}
  + \textcolor{BrickRed}{A_{\Psi}}\,\Psi_{\mathrm{i},\mathrm{j},\mathrm{k}}^\mathrm{n} \label{eq:zetaimp} \\ \nonumber
&\qquad\qquad\quad
  + \textcolor{BrickRed}{A_{\Phi'}}\,\Phi_{\mathrm{i},\mathrm{j},\mathrm{k}}^{\prime \mathrm{n}}
  + \textcolor{BrickRed}{A_{\nabla^2 \pi}}\,\nabla^2 \pi_{\mathrm{i},\mathrm{j},\mathrm{k}}^\mathrm{n}
  + \textcolor{BrickRed}{A_{\nabla^2 \Psi}}\,\nabla^2 \Psi_{\mathrm{i},\mathrm{j},\mathrm{k}}^\mathrm{n}
  + \textcolor{BrickRed}{A_{\delta P_m}}\,\delta P_{\mathrm{i},\mathrm{j},\mathrm{k}}^\mathrm{n}
\Bigr)
\Biggr] \, ,\\
\pi^{\mathrm{n}+1}
&=\frac{1}{1+\tfrac{\mathcal H\Delta\tau}{2}}
\left(\pi^{\mathrm{n}}
+\Delta\tau\big[\zeta^{\mathrm{n}+\frac12}+\Psi^{\mathrm{n}+\frac12}-\tfrac{\mathcal H}{2}\pi^{\mathrm{n}}\big]\right) \, .\label{eq:piimp}
\end{align}
\end{tcolorbox}
\label{tab:zeta_pi_imp}
\end{table}

\newpage
\subsection{Stress-energy tensor of dark energy}
We implement the $^0{}_0$ and $^i{}_j$ components of the dark energy stress–energy tensor on the lattice (the $^0{}_i$ term is redundant for implementation while is good to be implemented for consistency checks).
In \gev, each component carries an overall factor $a^3$ from the comoving-volume convention. Furthermore,
in the lattice projection of the $^i{}_j$ components, we eliminate the $\zeta'$ term from Eq. \eqref{eq:DE-Tmunu} by using the scalar field equation of motion, i.e. Eq. \eqref{eq:linearEOM}. Eqs. \eqref{eq:T00imp} and \eqref{eq:Tijimp} are implemented in \texttt{projection\_Tmunu\_kgb} function inside \texttt{gevolution.hpp} file.

\begin{table}[htbp]
    \centering
    \hspace{-0.5cm}
    \caption{Dark energy stress-energy tensor component implemented in the code.}
\begin{tcolorbox}[
  colframe=black,
  colback=white!10!white,
  boxrule=0.5mm,
  title=\textbf{Stress–Energy Tensor Projection},
  halign title=center,
  width=\textwidth
]
\begin{align}
\delta T^0_{0;\,\mathrm{i},\mathrm{j},\mathrm{k}}
&= a^3\Bigl\{
   \frac{M_{\mathrm{Pl}}^2}{a^2}\bigl[
    \alpha_{\mathrm{B}}\,\mathcal{H}\,\nabla^2\pi^n_{\mathrm{i},\mathrm{j},\mathrm{k}}
    +3\alpha_{\mathrm{B}}\,\mathcal{H}^2\,\Psi^n_{\mathrm{i},\mathrm{j},\mathrm{k}}
    -(3\alpha_{\mathrm{B}}+\alpha_{\mathrm{K}})\,\mathcal{H}^2\,\zeta^n_{\mathrm{i},\mathrm{j},\mathrm{k}} \label{eq:T00imp}
    \\
&\qquad    \nonumber+3\alpha_{\mathrm{B}}\,\mathcal{H}\,\Phi^{\prime\,n}_{\mathrm{i},\mathrm{j},\mathrm{k}}
    +\bigl(\alpha_{\mathrm{B}}\,\mathcal{H}'
      -\alpha_{\mathrm{B}}\,\mathcal{H}^2
      +\frac{a^2}{M_{\mathrm{Pl}}^2}(\bar{\rho}_\phi+\bar{P}_\phi)\bigr)
    \,3\,\mathcal{H}\,\pi^n_{\mathrm{i},\mathrm{j},\mathrm{k}}
  \bigr]\Bigr\},\\[8pt]
\delta T^p_{q;\,\mathrm{i},\mathrm{j},\mathrm{k}}
&= a^3\,\delta^p_{q}\Bigl[  
  \textcolor{BrickRed}{C_{\delta P_m}}\,\delta P^n_{m;\,\mathrm{i},\mathrm{j},\mathrm{k}}
  + \textcolor{BrickRed}{C_{\nabla^2\Psi}}\,\nabla^2\Psi^n_{\mathrm{i},\mathrm{j},\mathrm{k}}
  + \textcolor{BrickRed}{C_{\psi}}\,\Psi^n_{\mathrm{i},\mathrm{j},\mathrm{k}}
  + \textcolor{BrickRed}{C_{\phi'}}\,\Phi^{\prime\,n}_{\mathrm{i},\mathrm{j},\mathrm{k}}
  \label{eq:Tijimp} \\
&\quad \nonumber
  + \textcolor{BrickRed}{C_{\nabla^2\pi}}\,\nabla^2\pi^n_{\mathrm{i},\mathrm{j},\mathrm{k}}
  + \textcolor{BrickRed}{C_{\zeta}}\,\zeta^n_{\mathrm{i},\mathrm{j},\mathrm{k}}
  + \textcolor{BrickRed}{C_{\pi}}\,\pi^n_{\mathrm{i},\mathrm{j},\mathrm{k}}
  \Bigr].
\end{align}

where

\begin{align*}
\textcolor{BrickRed}{C_{\delta P_m}}
&= -\frac{3\,\alpha_{\mathrm{B}}^2}{3\,\alpha_{\mathrm{B}}^2 + 2\,\alpha_{\mathrm{K}}}\, ,\\[4pt]
\textcolor{BrickRed}{C_{\nabla^2\Psi}}
&= -\frac{2\,M_{\mathrm{Pl}}^2\,\alpha_{\mathrm{B}}^2}{(3\,\alpha_{\mathrm{B}}^2+2\,\alpha_{\mathrm{K}})\,a^2}\, ,\\[4pt]
\textcolor{BrickRed}{C_{\psi}}
&= \frac{6\,\alpha_{\mathrm{B}}}{(3\,\alpha_{\mathrm{B}}^2+2\,\alpha_{\mathrm{K}})\,a^2}
  \Bigl[
    M_{\mathrm{Pl}}^2\bigl(\alpha_{\mathrm{B}}\,\mathcal{H}^2
    -\alpha_{\mathrm{B}}'\,\mathcal{H}
    -\alpha_{\mathrm{B}}\,\mathcal{H}'\bigr)
    - a^2(\bar{\rho}_\phi+\bar{P}_\phi)
  \Bigr]\, ,\\[4pt]
\textcolor{BrickRed}{C_{\phi'}}
&= \frac{1}{\mathcal{H}}\;\textcolor{BrickRed}{C_{\psi}}\, ,\\[4pt]
\textcolor{BrickRed}{C_{\nabla^2\pi}}
&= -\frac{2\,\alpha_{\mathrm{B}}}{(3\,\alpha_{\mathrm{B}}^2+2\,\alpha_{\mathrm{K}})\,\mathcal{H}\,a^2}
  \Bigl[
    M_{\mathrm{Pl}}^2\bigl(\alpha_{\mathrm{B}}'\,\mathcal{H}
    +\alpha_{\mathrm{B}}\,\mathcal{H}'\bigr)
    + a^2(\bar{\rho}_\phi+\bar{P}_\phi)
  \Bigr]\, ,\\[4pt]
\textcolor{BrickRed}{C_{\zeta}}
&= -\frac{2}{(3\,\alpha_{\mathrm{B}}^2+2\,\alpha_{\mathrm{K}})\,a^2}
  \Bigl[
    M_{\mathrm{Pl}}^2\bigl(\alpha_{\mathrm{B}}\alpha_{\mathrm{K}}\,\mathcal{H}^2
    +\alpha_{\mathrm{K}}\,\mathcal{H}\,\alpha_{\mathrm{B}}'
    -\alpha_{\mathrm{B}}\,\mathcal{H}\,\alpha_{\mathrm{K}}'\bigr)
    - \alpha_{\mathrm{K}}\,a^2(\bar{\rho}_\phi+\bar{P}_\phi)
  \Bigr]\, ,\\[4pt]
\textcolor{BrickRed}{C_{\pi}}
&= \frac{2}{(3\,\alpha_{\mathrm{B}}^2+2\,\alpha_{\mathrm{K}})\,\mathcal{H}\,a^2}
  \Bigl[
    3\,M_{\mathrm{Pl}}^2\,\alpha_{\mathrm{B}}^2\bigl(3\,\mathcal{H}^2\,\mathcal{H}'
    - \mathcal{H}\,\mathcal{H}''- \mathcal{H}'^2\bigr)\\
&\quad    +3\,M_{\mathrm{Pl}}^2\,\alpha_{\mathrm{B}}\,\alpha_{\mathrm{B}}'\bigl(\mathcal{H}^3
    - \mathcal{H}\,\mathcal{H}'\bigr)
    +3\,\alpha_{\mathrm{B}}\,a^2\bigl(\mathcal{H}^2(\bar{\rho}_\phi+\bar{P}_\phi)
    - \mathcal{H}'(\bar{\rho}_\phi+\bar{P}_\phi)\bigr)\\
&\quad    + \alpha_{\mathrm{K}}\,\mathcal{H}\,P_{\phi}'\,a^2
  \Bigr].
\end{align*}
\end{tcolorbox}
    \label{tab:stress-energy_DE_imp}
\end{table}
\section{Gauge transformations}
\label{sec:GaugeTrans}
In \hiclass code, the KGB dark energy model is implemented in the field language and the scalar field perturbations $v_x$ and its time derivative $v^\prime_x$, as well as density contrast of other species are given in synchronous gauge. The purpose of this appendix is to relate the metric and matter perturbation in synchronous gauge (\hiclass default) to the conformal Newtonian gauge (\gev default).
To do the gauge transformation and bring everything to conformal Newtonian gauge we follow \cite{Ma:1995ey} and \cite{Hassani:2019lmy}.\\
\subsection{General gauge transformation rules for scalar perturbations}
We start with the general form of a perturbed FLRW metric, which can be written as
\begin{align}
g_{00} &= -a^2(\tau)\Bigl[1+2\Psi(\vec{x},\tau)\Bigr] \, ,\\[1mm]
g_{0i} &= a^2(\tau)\,w_i(\vec{x},\tau) \, ,\\[1mm]
g_{ij} &= a^2(\tau)\Bigl\{[1-2\Phi(\vec{x},\tau)]\delta_{ij}+\chi_{ij}(\vec{x},\tau)\Bigr\} \, ,\quad\text{with } \chi_{ii}=0  \, .
\label{eq:generalMet}
\end{align}
Here, $a(\tau)$ is the scale factor; $\Psi$ and $\Phi$ denote the scalar metric perturbations; $w_i$ represents the vector perturbations; and $\chi_{ij}$ is the traceless tensor perturbation.

As a first step, we will derive how the perturbations transform under an infinitesimal coordinate change of the form
\begin{equation}
  \hat{x}^\mu \rightarrow x^\mu + d^\mu(x^{\nu})   \, .
\end{equation}
We decompose the infinitesimal transformations $d^\mu$ into temporal and spatial scalar components, $d^\mu=(\,\alpha\,,\;\partial^i\beta)\,$. So
\begin{align}
\hat{x}^0 &= x^0 + \alpha(\tau,\vec{x}) \, ,\\
\hat{\vec{x}} &= \vec{x} + \vec{\nabla}\beta(\tau,\vec{x}) + \vec{\epsilon}(\tau,\vec{x}) \, , \qquad \vec{\nabla}.\vec{\epsilon} = 0 \,,
\end{align}
where the vector $\vec{\epsilon}$ is transverse and typically ignored for scalar perturbations. Thus, only the scalar gauge parameters $\alpha$ and $\beta$ are needed. \\
Under a change of coordinates, the metric transforms as
\begin{equation}
    \hat{g}_{\mu\nu}(\hat{x}) = g_{\alpha\beta}(x)\, \frac{\partial x^\alpha}{\partial \hat{x}^\mu}\, \frac{\partial x^\beta}{\partial \hat{x}^\nu}\,.
\end{equation}
The inverse transformation is given
\begin{equation}
  x^\mu = \hat{x}^\mu - d^\mu .  
\end{equation}
Taking the derivative with respect to $\hat{x}^\mu$, we get
\begin{equation}
\begin{aligned}
 \frac{\partial x^\alpha}{\partial \hat{x}^\mu} = \delta^\alpha_\mu - \partial_\mu d^\alpha \,,   \qquad
  \frac{\partial x^\beta}{\partial \hat{x}^\nu} = \delta^\beta_\nu - \partial_\nu d^\beta \,. 
\end{aligned}
\end{equation}
Therefore, we can write
\begin{align}
\hat{g}_{\mu\nu}(\hat{x}) &= g_{\alpha\beta}(x)\, \frac{\partial x^\alpha}{\partial \hat{x}^\mu}\, \frac{\partial x^\beta}{\partial \hat{x}^\nu}\,.\nonumber\\
& = g_{\alpha\beta}(x)\Big(\delta^\alpha_\mu - \partial_\mu d^\alpha\Big)\Big(\delta^\beta_\nu - \partial_\nu d^\beta\Big) \nonumber\\
& = g_{\mu\nu}(x) - g_{\alpha\nu}(x)\partial_\mu d^\alpha - g_{\mu\beta}(x)\partial_\nu d^\beta \, .
\label{eq:GT}
\end{align}

The metric in the new coordinates $g_{\mu\nu}(\hat{x})$ must be re-expressed in terms of $x$. Taylor expanding the metric $g_{\mu\nu}(\hat{x})$ about $x$
\begin{equation}
\hat{g}_{\mu\nu}(\hat{x}) = \hat{g}_{\mu\nu}(x + d) = \hat{g}_{\mu\nu}(x) + d^\gamma \partial_\gamma g_{\mu\nu}(x)\,.   
\end{equation}
Substituting this into Eq. \eqref{eq:GT}, we can write the expression to \textit{first order} as
\begin{equation}
 \hat{g}_{\mu \nu}(x)=g_{\mu \nu}(x)-g_{\mu \beta}(x) \partial_\nu d^\beta-g_{\alpha \nu}(x) \partial_\mu d^\alpha-d^\gamma \partial_\gamma g_{\mu \nu}(x) \, .   
 \label{eq:FGT}
\end{equation}\\
\textbf{00 component:}
\begin{align}
    \hat{g}_{00}(x) &= g_{00}(x) -2g_{00}(x)\partial_0d^0 -d^\gamma\partial_\gamma g_{00}\nonumber\\
    & = -a^2(1+2\Psi) -2a^2(1+2\Psi)\alpha^\prime - \alpha\partial_0\big[-a^2(1+2\Psi)\big]\nonumber\\
    & = -a^2(1+2\Psi) -2a^2\alpha^\prime - \alpha\big[-2aa^\prime(1+2\Psi)-2a^2\Psi^\prime\big]\nonumber\\
    & =  -a^2(1+2\Psi) -2a^2\alpha^\prime +\alpha2aa^\prime\nonumber\\
    & = -a^2\big[1+2\Psi-2\alpha^\prime - 2\mathcal{H}\alpha \big] \, ,
\end{align}
where we have assumed that $\alpha$ (or in general the coordinate transformation $d^\mu$) is at the same order of the metric perturbation.
By definition the transformed metric in the new coordinates is written in the same conformal Newtonian form
$
\hat{g}_{00}(x)=-a^2[1+2\hat{\Psi}]. 
$
Thus,
\begin{equation}
    -a^2[1+2\hat{\Psi}]\, = -a^2\big[1+2\Psi-2\alpha^\prime - 2\mathcal{H}\alpha \big] \, ,
\end{equation}
which leads to
\begin{equation}
    \hat{\Psi} = \Psi-\alpha^\prime - \mathcal{H}\alpha \, .
\end{equation}\\
\textbf{ij component (trace part):} Similarly for the spatial components ($\mu=i,\;\nu=j$), we have
\begin{equation}
  \hat{g}_{ij}(x)= g_{ij}(x) - g_{ik}(x)\,\partial_j d^k - g_{kj}(x)\,\partial_i d^k - d^0\,\partial_0 g_{ij}(x) - d^k\,\partial_k g_{ij}(x)\,.
  \label{eq:ijGT}
\end{equation}
Thus
\begin{align}
   \hat{g}_{ij}(x) &= a^2(1-2\Phi)\delta_{ij} - a^2(1-2\Phi)[\delta_{ik}\,\partial_j d^k + \delta_{kj}\,\partial_i d^k] - [d^0\,\partial_0  +d^k\,\partial_k] a^2(1-2\Phi)\delta_{ij}\,\nonumber\\
   & = a^2(1-2\Phi)\delta_{ij} - a^2(1-2\Phi)[\delta_{ik}\,\partial_j \partial^k\beta + \delta_{kj}\,\partial_i \partial^k\beta ] - [\alpha\,\partial_0  +\partial^k\beta \,\partial_k] a^2(1-2\Phi)\delta_{ij}\,\nonumber\\
   &=  a^2(1-2\Phi)\delta_{ij} - a^2(1-2\Phi)[\delta_{ik}\,\partial_j \partial^k\beta + \delta_{kj}\,\partial_i \partial^k\beta ] - 2\alpha aa^\prime\delta_{ij}\,\nonumber\\
   & = a^2(1-2\Phi)\delta_{ij} - 2a^2\partial_i\partial^j\beta - 2\alpha aa^\prime\delta_{ij}\, ,
   \label{eq:phiGT}
\end{align}
where we used  $d^k = \partial^k\beta$, and wrote
$
\partial_j d^k = \partial_j\partial^k \beta\,  
$.
Since we are interested in $\Phi$ we can take the trace of \eqref{eq:phiGT}
\begin{align}
    a^2(1-2\hat{\Phi}) &= \frac{\delta^{ij}}{3}\Big[ a^2(1-2\Phi)\delta_{ij} - 2a^2\partial_i\partial^j\beta - 2\alpha aa^\prime\delta_{ij}\,\Big]\nonumber\\
       &=  a^2(1-2\Phi) - \frac{2}{3}a^2\nabla^2\beta - 2\alpha aa^\prime\,\,\nonumber\\
      &=  a^2(1-2\Phi) - \frac{2}{3}a^2\nabla^2\beta - 2\alpha a^2\mathcal{H}\, .
\end{align}
Hence we will have 
\begin{equation}
\hat{\Phi} = \Phi +\frac{1}{3}\nabla^2\beta +\alpha\mathcal{H} \, .
\end{equation}\\
\textbf{0i component and ij component (traceless part):} We can apply the same procedure for the $0i$ component and the $ij$ traceless part, which results in
\begin{align}
\hat{w}_i(\vec{x}, \tau) & =w_i(\vec{x}, \tau)+\partial_i \alpha(\vec{x}, \tau)-\partial_i {\beta^\prime}(\vec{x}, \tau)-{\epsilon^\prime_i}(\vec{x}, \tau) \, , \\
\hat{\chi}_{i j}(\vec{x}, \tau) & =\chi_{i j}(\vec{x}, \tau)-2\left\{\left(\partial_i \partial_j-\frac{1}{3} \delta_{i j} \nabla^2\right) \beta(\vec{x}, \tau)+\frac{1}{2}\left(\partial_i \epsilon_j+\partial_j \epsilon_i\right)\right\} \, .
\end{align}
From the vector and traceless tensor component we can extract the scalar mode which are hidden in the longitudinal parts
\begin{align}
\hat{w}_{i}^{\|} &= w_{i}^{\|}+ \partial_i \alpha(\vec{x}, \tau)-\partial_i {\beta^\prime}(\vec{x}, \tau) \, ,\\
\hat{\chi}_{i j}^{\|}(\vec{x}, \tau) & =\chi_{i j}(\vec{x}, \tau)-2\left(\partial_i \partial_j-\frac{1}{3} \delta_{i j} \nabla^2\right) \beta(\vec{x}, \tau) \, .
\end{align}

In summary, considering the scalar modes only, we will have the following transformations
\begin{tcolorbox}[colframe=black,colback=white!75!brown,boxrule=0.5mm,title=\textbf{gauge transformations for scalar modes},halign title=center]
   \begin{align}
\hat{\Psi}(\vec{x}, \tau) & =\Psi(\vec{x}, \tau)-{\alpha^\prime}(\vec{x}, \tau)-\mathcal{H} \alpha(\vec{x}, \tau)  \, ,
\label{eq:PsiT}\\
\hat{w}_{i}^{\|} &= w_{i}^{\|}+ \partial_i \alpha(\vec{x}, \tau)-\partial_i {\beta^\prime}(\vec{x}, \tau)  \, ,
\label{eq:wGT}\\
\hat{\Phi}(\vec{x}, \tau) & =\Phi(\vec{x}, \tau)+\frac{1}{3} \nabla^2 \beta(\vec{x}, \tau)+\mathcal{H} \alpha(\vec{x}, \tau)  \, , 
\label{eq:PhiGT}\\
\hat{\chi}_{i j}^{\|}(\vec{x}, \tau) & =\chi_{i j}(\vec{x}, \tau)-2\left(\partial_i \partial_j-\frac{1}{3} \delta_{i j} \nabla^2\right) \beta(\vec{x}, \tau)  \, .
\label{eq:TenGT}
\end{align}
\end{tcolorbox}

\subsection{Metric perturbations}

\textbf{Synchronous gauge:}
In the synchronous gauge the metric is defined such that the components $g_{00}$ and $g_{0i}$ are unperturbed. The line element is written as 
\begin{equation}
ds^2 = a^2(\tau)\left[ -d\tau^2 + (\delta_{ij}+h_{ij})dx^i dx^j \right]  \, ,
\label{eq:syncMet}
\end{equation}
where $h_{ij}$ represents the perturbation to the spatial part of the metric which can be decomposed into a trace part and a traceless part. Specifically, one can write
\begin{equation}
    h_{ij} = \frac{h\delta_{ij}}{3} + h_{i j}^{\|}+h_{i j}^{\perp}+h_{i j}^T  \, ,
\end{equation}
where $h = h_{ii}$ is the trace, and the other terms include the traceless components. In this decomposition, the divergence of $h_{ij}^{\|}$ yields a purely longitudinal vector, while the divergence of $h_{ij}^{\perp}$ gives a transverse vector, and $h_{ij}^T$ is defined to be transverse. 
The scalar part of $h_{ij}$ contains two degrees of freedom. One is given by the trace $h$ and the other is contained in the longitudinal part $h_{i j}^{\|}$. Although $h_{ij}^{\parallel}$ is a vector, its longitudinal nature implies that it can be completely described by a single scalar field $\lambda$, such that
\begin{equation}
   h_{ij}^{\|} = \left(\partial_i \partial_j - \frac{1}{3}\delta_{ij}\nabla^2\right)\lambda  \, .
\end{equation}
Therefore, the scalar mode of $h_{ij}$ in Fourier space, can be written as 
\begin{equation}
h_{ij}(\vec{x},\tau)=\int d^3k\, e^{i\vec{k}\cdot\vec{x}}\left\{\hat{k}_i\hat{k}_j\, h(\vec{k},\tau)+\left(\hat{k}_i\hat{k}_j-\frac{1}{3}\delta_{ij}\right) 6\eta(\vec{k},\tau)\right\}, \qquad \vec{k} = k\hat{k} \, ,
\end{equation}
where we used $   6\,\eta(\vec k,\tau)\;\equiv\;-\,k^2\,\lambda(\vec k,\tau).$ Considering only $h_{ij}^{\|}(\vec{x},\tau)$
\begin{equation}
    h_{ij}^{\|}(\vec{x},\tau)=\int d^3k\, e^{i\vec{k}\cdot\vec{x}}\, \left(\hat{k}_i\hat{k}_j-\frac{1}{3}\delta_{ij}\right) \left\{h(\vec{k},\tau)+6\eta(\vec{k},\tau)\right\}, 
    \label{eq:Longhij}
\end{equation}
where we introduced the fields $h(\vec{k},\tau)$ and $\eta(\vec{k},\tau)$. We will use \eqref{eq:Longhij} later.
\vspace{1.0em}
\\
 \textbf{Newtonian gauge:} In the conformal Newtonian gauge, the line element takes the form as
\begin{equation}
ds^2 = a^2(\tau)\left[ -(1+2\Psi)d\tau^2 + (1-2\Phi)dx^i dx_i \right]  \, .
\end{equation}
Now the goal is to relate the scalar metric perturbation $(\Psi, \Phi)$ in conformal Newtonian gauge to the scalar modes $(h,\eta)$ in synchronous gauge\footnote{Note that here we only consider the scalar part, since at linear order no vector or tensor modes are generated and, if present, they are gauge-invariant on an FLRW background according to the Stewart–Walker lemma.}. We use $\hat{x}^\mu$ to denote the synchronous gauge and $x^\mu$ to represent the conformal Newtonian gauge. Since $\hat{\Psi}$ is not present in the synchronous gauge, from \eqref{eq:PsiT} we can deduce
\begin{equation}
\Psi(\vec{x}, \tau) = \alpha^\prime(\vec{x}, \tau)  +\mathcal{H} \alpha(\vec{x}, \tau)     \, .
\end{equation}
We also know that there is no vector perturbations (i.e. $\hat{w}_{i}^{\|} = w_{i}^{\|}$ =0) in both gauges, therefore from \eqref{eq:wGT} we can write
\begin{equation}
 \alpha(\vec{x}, \tau) =  {\beta^\prime}(\vec{x}, \tau)   \, . 
 \label{eq:alpha_betaprime}
\end{equation}
Regarding \eqref{eq:PhiGT}, $\hat{\Phi}$ in synchronous gauge is given by $-\frac{1}{6}h(\vec{x},\tau)$ (to check you can plug this into \eqref{eq:generalMet} and compare it to \eqref{eq:syncMet}), hence
\begin{equation}
    \Phi(\vec{x}, \tau) = -\frac{1}{6}h(\vec{x},\tau) -\frac{1}{3} \nabla^2 \beta(\vec{x}, \tau)-\mathcal{H} \alpha(\vec{x}, \tau) \, .
\end{equation}
Finally, considering \eqref{eq:TenGT}, we notice that $\chi_{ij} = 0$ in conformal Newtonian gauge and in synchronous gauge we have $\hat{\chi}^{\|}_{ij} = h^{\|}_{ij}$. 
In Fourier space, \eqref{eq:TenGT} takes the form as 
\begin{equation}
    \hat{\chi}_{i j}^{\|}(\vec{x}, \tau)  =\int d^3k\, e^{i\vec{k}\cdot\vec{x}}\left(\hat{k}_i \hat{k}_j-\frac{1}{3} \delta_{i j}\right) k^2\beta(\vec{x}, \tau) \, .
    \label{eq:Longchiij}
\end{equation}
Note that we used
\begin{equation}
    \partial_i \partial_j \longrightarrow -k_ik_j = k^2\hat{k}_i \hat{k}_j~ , \qquad \nabla^2 = -k^2 \,,
\end{equation}
therefore, comparing \eqref{eq:Longhij} and \eqref{eq:Longchiij}, and noting \eqref{eq:alpha_betaprime} we can deduce the following expressions for $\beta$ and $\alpha$
\begin{equation}
    \beta = \frac{h(\vec{k},\tau)+6\eta(\vec{k},\tau)}{2k^2}~,\qquad \alpha = \frac{h^{\prime}(\vec{k},\tau)+6\eta^{\prime}(\vec{k},\tau)}{2k^2} \, .
\end{equation}
We can summarize the relation between the scalar metric perturbations in two different gauges as follow
\begin{tcolorbox}[colframe=black,colback=white!75!brown,boxrule=0.5mm,title=\textbf{$(\Psi,\Phi) \Longrightarrow (h,\eta)$},halign title=center]
   \begin{align}
\Psi(\vec{x}, \tau) &= \alpha^\prime(\vec{x}, \tau)  +\mathcal{H} \alpha(\vec{x}, \tau)     \, ,
\\
    \Phi(\vec{x}, \tau) &= -\frac{1}{6}h(\vec{x},\tau) -\frac{1}{3} \nabla^2 \beta(\vec{x}, \tau)-\mathcal{H} \alpha(\vec{x}, \tau)  \, .
\label{eq:finalPhiGT}
\end{align}
\end{tcolorbox}
\subsection{Matter perturbations}

Now, we can apply the same analogy to derive the relation for $\delta T^\mu_{~\nu}$ in two gauges. At the linear order in perturbation, the stress-energy tensor is given by
\begin{align}
T^0_{~0} & =-(\bar{\rho}+\delta \rho) \, ,\\
T^0_{~i} & =(\bar{\rho}+\bar{P}) v_i=-T^i_{~0}\, , \\
T^i_{~j} & =(\bar{P}+\delta P) \delta_j^i+\Sigma_j^i \, , \quad \Sigma_i^i=0\, ,
\end{align}\\
The transformation for stress-energy tensor $T^{\mu}_{~\nu}$  under the small coordinate change
\begin{equation}
 \hat{x}^{\mu} \;=\; x^{\mu} \;+\;d^{\mu}(x)\, ,   
\end{equation}
reads
\begin{equation}
\hat{T}^{\mu}_{~\nu}(\hat{x}) = \frac{\partial \hat{x}^\mu}{\partial x^\alpha}\, \frac{\partial x^\beta}{\partial \hat{x}^\nu}T^{\alpha}_{~\beta}(x)\,.
\end{equation}
Taking the derivatives yields
\begin{equation}
\begin{aligned}
 \frac{\partial \hat{x}^\mu}{\partial x^\alpha} &= \delta^\mu_\alpha + \partial_\alpha d^\mu \,,   \\
  \frac{\partial x^\beta}{\partial \hat{x}^\nu} &= \delta^\beta_\nu - \partial_\nu d^\beta \,. 
\end{aligned}
\end{equation}
Therefore
\begin{align}
\hat{T}^{\mu}_{~\nu}(\hat{x}) &= \frac{\partial \hat{x}^\mu}{\partial x^\alpha}\, \frac{\partial x^\beta}{\partial \hat{x}^\nu}T^{\alpha}_{~\beta}(x) \nonumber\\
& = \Big(\delta^\mu_\alpha + \partial_\alpha d^\mu  \Big)\Big(\delta^\beta_\nu - \partial_\nu d^\beta\Big) T^{\alpha}_{~\beta}(x)\nonumber\\
& = T^{\mu}_{~\nu}(x) -\partial_\nu d^\beta T^{\mu}_{~\beta}(x) +\partial_\alpha d^\mu T^{\alpha}_{~\nu}(x) \, .
\label{eq:TensGT}
\end{align}

At the same spacetime coordinate values, expanding to \emph{first order} in the small gauge function \(d^{\mu}\) gives a formula completely analogous to the metric
\begin{equation}
 \hat{T}^{\mu}{}_{\nu}(x)=
 T^{\mu}_{~\nu}(x) -\partial_\nu d^\beta T^{\mu}_{~\beta}(x) +\partial_\alpha d^\mu T^{\alpha}_{~\nu}(x) -
d^{\gamma}\,\partial_{\gamma}T^{\mu}{}_{\nu}(x)\, .
\end{equation}
\textbf{00 component:} For 00 component the above equation reads
\begin{align}
\hat{T}^{0}{}_{0}(x)&=
T^{0}{}_{0}(x)-\,\partial_{0}d^{\beta}T^{0}{}_{\beta}(x)+
\partial_{\alpha}d^{0}T^{\alpha}{}_{0}(x)-
d^{\gamma}\,\partial_{\gamma}T^{0}{}_{0}(x)\nonumber \\
&=T^{0}{}_{0}(x)-\,\partial_{0}d^{0}T^{0}{}_{0}(x)+
\partial_{0}d^{0}T^{0}{}_{0}(x)-
d^{0}\,\partial_{0}T^{0}{}_{0}(x)\nonumber\\
& = T^{0}{}_{0}(x)-\alpha\big(-{\bar{\rho}}^{\prime}-\delta\rho\big)\\
& = T^{0}{}_{0}(x)+\alpha{\bar{\rho}}^{\prime}\, .
\end{align}
Because the background fluid is at rest, $T^0{}_i$ and $T^i{}_0$ vanish at zeroth order and are themselves first‐order quantities.  Consequently, any term involving these (whether multiplied by a gauge function like $\alpha$ or by its derivative $\partial_\mu d^\nu$) is second order in perturbations and is dropped in the linearized analysis. Also, $d^{\gamma}\,\partial_{\gamma}T^{0}{}_{0}$ is first order if we keep the background part of $T^{0}{}_{0}$.\\
By definition
\begin{equation}
 T^0{}_{0}=-\,(\bar\rho+\delta\rho)
\quad\Longrightarrow\quad
\delta\rho=-\,T^0{}_{0}-\bar\rho \, .   
\end{equation}
Since $\hat T^0{}_{0}=T^0{}_{0}+\alpha\,\bar\rho'$, it follows that
 \begin{equation}
\delta\rho_{\rm sync}
=-\,\hat T^0{}_{0}-\bar\rho
=-\,\bigl[T^0{}_{0}+\alpha\,\bar\rho'\bigr]-\bar\rho
=\delta\rho_{\rm newt}-\alpha\,\bar\rho' \, ,   
 \end{equation}
and finally
\begin{equation}
 \delta^{\rm Sync}
= \delta^{\rm Newt}
  -\frac{\bar\rho'}{\bar\rho}\alpha \, ,   
\end{equation}\\
\textbf{0i component:} Setting $\mu=0$, $\nu=i$, one obtains
\begin{equation}
\hat T^{0}{}_{i}
= T^{0}{}_{i}
- \partial_i d^\beta\,T^{0}{}_{\beta}
+ \partial_\alpha d^0\,T^{\alpha}{}_{i}
- d^\gamma\,\partial_\gamma T^{0}{}_{i}
\;+\;\mathcal O(d^2) \, .
\end{equation}
considering the term $\partial_i d^\beta\,T^{0}{}_{\beta}$, only $\beta = 0$ contributes at zeroth order
\begin{equation}
     -\partial_i d^0\,T^0{}_0
 = -(\partial_i\alpha)\bigl[-(\bar\rho+\delta\rho)\bigr]
 \;\approx\;+\,\bar\rho\,\partial_i\alpha
 \quad(\alpha\,\delta\rho',\,\partial_i\alpha\,\delta\rho \text{ dropped}) \, .
\end{equation}
In $\partial_\alpha d^0\,T^\alpha{}_i$ term, only $\alpha=j$ picks up the background spatial pressure
\begin{equation}
     \partial_j d^0\,T^j{}_i
 = (\partial_i\alpha)\,\bar P
 \quad(\text{since }T^j{}_i\big|_{\rm bg} = \bar P\,\delta^j_i) \, .
\end{equation}
Finally, the last term  vanishes at linear order because $\bar T^0{}_i=0$, and any $\delta T^0{}_i$ derivative times $d^\gamma$ is second order. Putting these together gives
\begin{equation}
\hat T^{0}{}_{i}
= T^{0}{}_{i} + (\bar\rho+\bar P)\,\partial_i\alpha \, ,
\end{equation}
dividing both sides by $(\bar\rho + \bar P)$ we get
\begin{equation}
 \hat v_i
 = v_i \;-\;\partial_i\alpha \, ,    
\end{equation}
or in Fourier space 
\begin{equation}
         \hat v_i(\vec k)
     = v_i(\vec k)
     - i\,k_i\,\alpha(\vec k) \, .
\end{equation}
By definition we have $\theta \equiv ik^iv_i \,$, thus 
\begin{align}
      \hat\theta(\vec k)
     &= ik^i\hat v_i(\vec k)\nonumber\\
     &= ik^iv_i(\vec k)
       - ik^i\bigl[ik_i\alpha(\vec k)\bigr]\nonumber\\
     &= \theta(\vec k)
       -k^2\alpha(\vec k) \, .   
\end{align}
Finally we can write
\begin{equation}
    \theta^{\rm Sync}(\vec k,\tau)
= \theta^{\rm Newt}(\vec k,\tau)
-k^2\alpha(\vec k,\tau) \, .
\end{equation}
\textbf{scalar field perturbations:} Considering scalar field perturbation $\pi$, we have 
 \begin{equation}
     \phi(x^\mu) = \bar{\phi}(x^\mu)+\delta\phi \, ,
 \end{equation}
where in unitary gauge $\bar{\phi}(x^\mu)$ coincides with time, i.e.
$\bar{\phi}(x^\mu) = x^0$ and therefore we have
\begin{equation}
  \phi(x^\mu) = x^0+\pi(x^\mu) \quad\Longrightarrow\quad
\partial_\mu\phi
=\delta^0{}_\mu + \partial_\mu\pi  \, .
\end{equation}
Under a diffeomorphism $x^\mu\mapsto\hat x^\mu$, a scalar field does not change its value
$$
\hat{\phi}(\hat{x}^\mu) = \phi(x^\mu)  \, .
$$
Thus at the same coordinate value
\begin{align}
\hat{\phi}(x^\mu) = \hat{\phi}(\hat{x}^\mu-d^\mu) &= \hat{\phi}(\hat{x}^\mu) - d^\mu\partial_\mu\hat{\phi}(\hat{x}^\mu)\nonumber\\
&= \phi(x^\mu) - d^\mu\partial_\mu\phi(x^\mu)\nonumber\\
&= \phi(x^\mu)-d^0\,(1+\pi') -d^i\,\partial_i\pi \nonumber\\
& = \phi(x^\mu)-\alpha  \, ,
\end{align}
Identifying $\hat x^\mu$ as synchronous and $x^\mu$ as Newtonian, and using $\hat\phi=\hat x^0+\hat\pi$ with $\hat x^0=x^0$ to first order, we relate the perturbations
\begin{equation}
    \pi^\text{Sync} = \pi^{\text{Newt}} -\alpha  \, ,
\end{equation} 
and equivalently
\begin{equation}
    \pi'^{\rm Sync} = \pi'^{\text{Newt}} -\alpha'  \, ,
\end{equation} 
which will be used for 
\begin{equation}
    \zeta^{\rm Newt} = \mathcal{H}\pi^{\rm Newt} + \pi'^{\rm Newt} -\Psi  \, .
\end{equation}
We can summarize our findings in this appendix as follow
\begin{tcolorbox}[
  colframe=black,colback=white!75!brown,boxrule=0.5mm,
  title=\textbf{Important equations},halign title=center,
  halign=center 
]
\begin{minipage}{0.8\linewidth} 
\begin{align}
\beta(\vec{k},\tau) &= \frac{h(\vec{k},\tau)+6\eta(\vec{k},\tau)}{2k^2} \, , \\
\alpha(\vec{k},\tau) &= \frac{h'(\vec{k},\tau)+6\eta'(\vec{k},\tau)}{2k^2} \, , \\
\Psi(\vec{k},\tau) &= \alpha'(\vec{k},\tau)+\mathcal{H}\,\alpha(\vec{k},\tau) \, , \\
\Phi(\vec{k},\tau) &= \eta(\vec{k},\tau)-\mathcal{H}\,\alpha(\vec{k},\tau) \, , \\
\delta^{\text{Sync}}(\vec{k},\tau) &= \delta^{\text{Newt}}(\vec{k},\tau)
 - \alpha(\vec{k},\tau)\frac{\bar{\rho}'}{\bar{\rho}}, \label{eq:densGT}\\
\theta^{\text{Sync}}(\vec{k},\tau) &= \theta^{\text{Newt}}(\vec{k},\tau)
 - k^2 \alpha(\vec{k},\tau) \, , \label{eq:thetGT}\\
\pi^{\text{Sync}}(\vec{k},\tau) &= \pi^{\text{Newt}}(\vec{k},\tau)
 - \alpha(\vec{k},\tau) \, , \label{eq:piGT}\\
\zeta^{\text{Newt}}(\vec{k},\tau) &= \mathcal{H}\,\pi^{\text{Newt}}(\vec{k},\tau)
 + (\pi^{\text{Newt}})'(\vec{k},\tau) - \Psi(\vec{k},\tau) \, .\label{eq:zetaGT}
\end{align}
\label{tab:GT}
\end{minipage}
\end{tcolorbox}

\subsection{Initial condition from \hiclass}
When running \kgb, all the perturbation are provided with \hiclass at $z = 100$ in synchronous gauge and gauge transformed accordingly via expressions presented in Table \ref{tab:GT}. These are done in the header file named \texttt{hiclass\_tools.hpp} inside a function named \texttt{loadTransferFunctions}. Note that we can rewrite Eq. \eqref{eq:densGT} using the background equation

\begin{equation}
3 \mathcal{H} \bar{\rho}  + 3 \mathcal{H} \bar{P} + \rho^{\prime} = 0 \quad \Longrightarrow \rho^\prime = -3\mathcal{H}(\bar{\rho}+ \bar{P}) \, .
\end{equation}
Thus Eq. \eqref{eq:densGT} becomes
\begin{equation}
 \delta^{\text{Sync}}(\vec{k}, \tau) =  \delta^{\text{Newt}}(\vec{k}, \tau) + 3\mathcal{H}\alpha(\vec{k}, \tau)\big(1+w\big)  \,.
 \label{eq:densGTgev}
\end{equation}
We actually use Eq. \eqref{eq:densGTgev} to perform the gauge transformation for cold dark matter and baryon species with $w = 0$ and photon and ultra-relativistic species with $w = 1/3$. For the rest, including $\theta$, $\pi$ and $\zeta$ we use the Eqs. \eqref{eq:thetGT} to \eqref{eq:zetaGT}.

\section{Density contrast of dark energy }
\label{sec:deltaDE}
In our approach, we have kept the geometric side of the Einstein equations unchanged and simply add the KGB stress–energy tensor to the matter side as an extra source term.  In contrast, \hiclass reformulates both the geometric (left-hand) and stress–energy (right-hand) sides of the field equations—even for minimally coupled models like KGB—by absorbing modifications into an effective gravitational sector.  As a result, \hiclass does not explicitly output dark energy density perturbations in its transfer function; instead, one must reconstruct the dark energy density contrast manually as per Eq. \eqref{eq:deltarhoDE}. In Fourier space the equation becomes
\begin{align}
\delta\rho_{\rm DE} &=  \frac{M_\textrm{Pl}{}^2}{a^2} \Big(-\alpha_\textrm{B} \mathcal{H}  k^2\pi  + 3 \alpha_\textrm{B} \mathcal{H} ^2 \Psi  - (3 \alpha_\textrm{B}  + \alpha_\textrm{K}) \mathcal{H} ^2 \zeta  + 3 \alpha_\textrm{B} \mathcal{H}  \Phi ^{\prime}\Big)  \nonumber \\
&+ \left(  \frac{M_\textrm{Pl}{}^2}{a^2}\alpha_\textrm{B} \mathcal{H} ^{\prime}-  \frac{M_\textrm{Pl}{}^2}{a^2} \alpha_\textrm{B} \mathcal{H} ^2   + (\bar{\rho}_\phi  + \bar{P}_\phi )\right)3 \mathcal{H} \pi  \, .
\label{eq:delta_rho}
\end{align}
Note that Eq. \eqref{eq:delta_rho} is written in the Newtonian gauge, therefore if we want to construct this in \hiclass we have to bring all the perturbations, $\pi$, $\pi'$ and $\zeta$ into the Newtonian 
gauge using the expression in Table \ref{tab:GT}, i.e. write
\begin{itemize}
    \item $\pi^{\rm Newt} = \pi^{\rm Sync} + \alpha  \Longrightarrow \pi^{\rm Newt} =  -v_{\rm smg }+ \alpha $  \, ,
    \item  $\pi'^{\rm Newt} = \pi'^{\rm Sync} + \alpha'  \Longrightarrow \pi'^{\rm Newt} = -v'_{\rm smg } + \Psi - \mathcal{H}\alpha$\,,
    \item $\zeta^{\rm Newt} = \mathcal{H}\pi^{\rm Newt} + \pi'^{\rm Newt} - \Psi  $  \, ,
\end{itemize}
where we used the fact that $\pi_{\rm KGB} = -v_{\rm smg}$.
The density contrast is then simply obtained by $\delta_{\rm DE}  = \delta \rho_{\rm DE}/ \bar{\rho} $. Finally, if someone decides to write this in the synchronous gauge, we can make use of Eq. \eqref{eq:densGT}.

 \bibliographystyle{JHEP}
\bibliography{biblio.bib}
\end{document}